\documentclass{article}

\PassOptionsToPackage{numbers, compress}{natbib}

\usepackage[eandd, preprint]{neurips_2026}

\usepackage[utf8]{inputenc} 
\usepackage[T1]{fontenc}    
\usepackage{url}            
\usepackage{booktabs}       
\usepackage{amsfonts}       
\usepackage{nicefrac}       
\usepackage{microtype}      
\usepackage{xcolor}         

\usepackage{amsmath}
\usepackage{amssymb}
\usepackage{mathtools}
\usepackage{amsthm}
\usepackage{longtable}
\usepackage{siunitx}
\usepackage{subcaption}
\usepackage{wrapfig}
\usepackage[table]{xcolor}
\usepackage{placeins}
\usepackage{float}
\usepackage{tabularx}

\theoremstyle{plain}

\theoremstyle{definition}

\theoremstyle{remark}

\usepackage[textsize=tiny]{todonotes}

\usepackage[most]{tcolorbox}
\usepackage{listings}

\lstdefinestyle{promptstyle}{
  basicstyle=\ttfamily\footnotesize,
  columns=fullflexible,
  keepspaces=true,
  breaklines=true,
  breakatwhitespace=false,
  showstringspaces=false
}

\newtcblisting{promptbox}[1][]{%
  enhanced,
  breakable,
  colback=white,
  colframe=black!80,
  coltitle=white,
  fonttitle=\bfseries,
  colbacktitle=black!80,
  boxrule=1pt,
  arc=3pt,
  left=8pt,right=8pt,top=6pt,bottom=6pt,
  listing only,
  listing options={style=promptstyle},
  title={#1}
}

\definecolor{codebg}{HTML}{DAE3F5}
\definecolor{kwred}{HTML}{DD4F6D}
\definecolor{emgreen}{HTML}{75A21F}
\definecolor{numvio}{HTML}{C07FA0}
\definecolor{framegray}{HTML}{C9D3E6}

\lstdefinestyle{myscreenshot}{
  backgroundcolor=\color{codebg},
  basicstyle=\ttfamily\small,
  numbers=left,
  numberstyle=\ttfamily\scriptsize\color{black!45},
  stepnumber=1,
  numbersep=10pt,
  xleftmargin=2.6em,
  frame=single,
  rulecolor=\color{framegray},
  framesep=6pt,
  framerule=0.6pt,
  breaklines=true,
  breakatwhitespace=true,
  tabsize=2,
  showstringspaces=false,
  keywordstyle=\bfseries\color{kwred},
  commentstyle=\color{black!45},
  stringstyle=\color{numvio},
  emphstyle=\bfseries\color{emgreen},
}

\lstdefinestyle{myscreenshot_small}{
  style=myscreenshot,
  basicstyle=\ttfamily\scriptsize,
  numbers=none,
  xleftmargin=0em
}
\usepackage{xspace}

\usepackage[
  pagebackref=false,
  breaklinks=true,
  colorlinks=true,
  linkcolor=blue,
  citecolor=blue,
  urlcolor=blue
]{hyperref}

\usepackage[capitalize,noabbrev]{cleveref}

\newcommand{\HPCBench}{HPC-Bench\xspace}

\usepackage{xcolor}
\usepackage{xargs}      
\usepackage{ifthen}
\usepackage{marginnote} 
\usepackage{xparse}     
\usepackage[normalem]{ulem} 

\newif\ifdraft
\draftfalse

\newcounter{todocnt}
\newcounter{notecnt}
\newcounter{fixcnt}
\newcounter{citecnt}
\newcounter{expcnt}
\newcounter{donecnt}

\newcommand{\mnote}[2]{%
  \marginnote{\scriptsize\color{#1}\textbf{#2}}%
}

\newcommand{\itag}[2]{%
  {\color{#1}\textbf{[#2]}}%
}

\NewDocumentCommand{\MARK}{m m m}{%
  \ifdraft
    \itag{#1}{#2: #3}%
  \fi
}

\NewDocumentCommand{\MMARK}{m m}{%
  \ifdraft
    \mnote{#1}{#2}%
  \fi
}

\NewDocumentCommand{\TODO}{O{} m}{%
  \ifdraft
    \refstepcounter{todocnt}%
    \def\theid{TODO\#\arabic{todocnt}}%
    \MARK{red}{\theid}{#2}%
    \ifthenelse{\equal{#1}{m}}{\MMARK{red}{\theid}}{}%
  \fi
}

\NewDocumentCommand{\NOTE}{O{} m}{%
  \ifdraft
    \refstepcounter{notecnt}%
    \def\theid{NOTE\#\arabic{notecnt}}%
    \MARK{blue}{\theid}{#2}%
    \ifthenelse{\equal{#1}{m}}{\MMARK{blue}{\theid}}{}%
  \fi
}

\NewDocumentCommand{\FIX}{O{} m}{%
  \ifdraft
    \refstepcounter{fixcnt}%
    \def\theid{FIX\#\arabic{fixcnt}}%
    \MARK{orange}{\theid}{#2}%
    \ifthenelse{\equal{#1}{m}}{\MMARK{orange}{\theid}}{}%
  \fi
}

\NewDocumentCommand{\CITE}{O{} m}{%
  \ifdraft
    \refstepcounter{citecnt}%
    \def\theid{CITE\#\arabic{citecnt}}%
    \MARK{purple}{\theid}{#2}%
    \ifthenelse{\equal{#1}{m}}{\MMARK{purple}{\theid}}{}%
  \fi
}

\NewDocumentCommand{\EXP}{O{} m}{%
  \ifdraft
    \refstepcounter{expcnt}%
    \def\theid{EXP\#\arabic{expcnt}}%
    \MARK{teal}{\theid}{#2}%
    \ifthenelse{\equal{#1}{m}}{\MMARK{teal}{\theid}}{}%
  \fi
}

\NewDocumentCommand{\DONE}{O{} m}{%
  \ifdraft
    \refstepcounter{donecnt}%
    \def\theid{DONE\#\arabic{donecnt}}%
    {\color{green!50!black}\textbf{[\theid\ (done): #2]}}%
    \ifthenelse{\equal{#1}{m}}{\MMARK{green!50!black}{\theid}}{}%
  \fi
}

\NewDocumentCommand{\TODOA}{m O{} m}{%
  \TODO[#2]{#1: #3}%
}
\NewDocumentCommand{\NOTEH}{m O{} m}{%
  \NOTE[#2]{#1: #3}%
}
\NewDocumentCommand{\FIXA}{m O{} m}{%
  \FIX[#2]{#1: #3}%
}
\NewDocumentCommand{\CITEA}{m O{} m}{%
  \CITE[#2]{#1: #3}%
}
\NewDocumentCommand{\EXPA}{m O{} m}{%
  \EXP[#2]{#1: #3}%
}

\title{HPC-Bench: A Comprehensive Benchmark for High Performance Computing Codes}

\author{%
  Bowen Cui \\
  George Mason University \\
  \texttt{bcui2@gmu.edu}
  \And
  Junyu Yin \\
  George Mason University \\
  \texttt{jyin7@gmu.edu}
  \And
  Tejas Ramesh \\
  George Mason University  \\
  \texttt{tramesh2@gmu.edu}
  \AND
  Leo Lim \\
  TJHSS  \\
  \texttt{reverileo@gmail.com}
  \And
  Oscar Hernandez \\
  Oak Ridge National Lab \\
  \texttt{oscar@ornl.gov}
  \And
  Keren Zhou \\
  George Mason University  \\
  \texttt{kzhou6@gmu.edu}
}

\begin{document}

\maketitle

\begin{abstract}
Large Language Models (LLMs) have shown strong capabilities in code generation and reasoning, motivating extensive benchmarks for evaluating functional correctness.
However, performance optimization in High-Performance Computing (HPC) poses fundamentally different challenges that are not captured by existing evaluations: HPC performance depends on low-level code, intricate memory access patterns, and computational kernels embedded within sophisticated execution contexts.
We introduce \HPCBench, a performance-grounded benchmark for evaluating LLMs as optimizers of computational kernels in realistic HPC workloads, including $114$ workloads across $16$ HPC computational motifs.
\HPCBench provides a fully automated, end-to-end evaluation pipeline that combines program-level correctness verification with two correctness-gated metrics, $\mathrm{Fast}@k$ and $\mathrm{Speedup}@k$, to characterize reliability and speedup under a fixed sampling budget.
Our evaluation of three representative LLMs across serial CPU, OpenMP, and CUDA settings shows that stricter speedup requirements substantially reduce the probability of finding a correct and fast optimization.
\HPCBench is available at \url{https://github.com/Deep-Learning-Profiling-Tools/HPC-Bench}.
\end{abstract}

\section{Introduction}
\label{sec:introduction}



Large Language Models (LLMs) have become widely used for code generation, completion, and reasoning, motivating extensive benchmarks for evaluating their programming capabilities~\cite{hossain2024togll,hou2024large,hui2024qwen2,dong2025survey}. 
Existing benchmarks span tasks from single-function code generation, such as HumanEval~\cite{chen2021evaluating}, to repository-scale development tasks, such as CoderEval~\cite{codereval}, HumanEvo~\cite{humanevo}, and RepoEval~\cite{repocoder}. 
However, these evaluations primarily measure functional correctness, while runtime efficiency is often treated as secondary or omitted entirely.

While recent code benchmarks have begun to target efficiency, they are limited either in scale (e.g., EffiBench~\cite{effibench} and EvalPerf~\cite{evalperf}) or to narrowly defined, well-structured problems (e.g., KernelBench~\cite{kernelbench} and TritonBench~\cite{tritonbench}).
Performance optimization in High-Performance Computing (HPC) code is a different challenge compared to these tasks.
HPC efficiency is shaped by a complex interplay of low-level code (e.g., C/C++, FORTRAN), various programming models (e.g., OpenMP, CUDA), memory access patterns, and sophisticated execution contexts, where the main computational kernels are deeply embedded within their surrounding code context.
Even minor code modifications, such as loop reordering~\cite{10.1145/143103.143132}, can induce substantial runtime differences. 

There have been some preliminary studies on benchmarking LLMs for HPC tasks, such as evaluations of DeepSeek and GPT-4 on classic solvers~\cite{deepseek_hpc_benchmark}, assessments of OpenAI Codex on GEMM and SpMV kernels~\cite{codex_hpc_eval}, and studies on MPI+OpenMP code generation~\cite{llm_hpcpp_eval}. 
However, these studies typically focus on only a narrow set of HPC codes and lack a standardized, performance-grounded benchmarking pipeline.

To fill this gap, we introduce \HPCBench, a performance-grounded benchmark for evaluating LLMs as local optimizers of profiling-identified computational kernels in realistic HPC programs.
We focus on local kernel rewriting rather than whole application rewriting, with the goal of evaluating standalone LLMs' optimization capability under a fixed sampling budget rather than evaluating agent-based workflows.

\paragraph{Challenges and Solutions.} Building HPC-Bench involves four main challenges: 
(i) Hotspots are deeply embedded in large HPC applications and are context- and input-dependent;
(ii) HPC applications span diverse computational motifs; 
(iii) Optimization effectiveness depends on both the programming model and hardware;
(iv) Correctness and performance are coupled. 

To address the above challenges, we make the following contributions in this paper:
\begin{itemize}
    \item \textbf{Evaluation Framework.} We identify computation hotspots through profiler-guided selection and manual filtering, and we recover the inputs and context necessary to reproduce performance hotspots in real HPC applications.
    \item \textbf{Suite.} We curate a comprehensive benchmark of $114$ workloads across $16$ computational motifs with standardized inputs, builds, and automated program-level correctness checking.
    \item \textbf{Experiments.} We benchmark representative LLMs under serial CPU, OpenMP, and CUDA optimization settings and analyze their strengths and failure modes across motifs, structural complexity stratum, and sampling budgets.
    \item \textbf{Metrics and Reporting}. We report sampling-aware, correctness-gated metrics (e.g., $\mathrm{Fast}@k$ and $\mathrm{Speedup}@k$) to capture the reliability-speedup tradeoff under a fixed sampling budget.
\end{itemize}

In our experiments, we identified that LLM-based HPC optimization is limited by the rapidly diminishing probability of producing correct optimizations as task structural complexity and speedup requirements increase. 
This trend manifests consistently across CPUs and GPUs, while its severity varies with computational motifs and benchmark structural complexity.

\section{Related Work}

\paragraph{Code Benchmarks.}

The rapid advancements in the capabilities of LLMs have motivated extensive efforts to systematically evaluate their coding abilities, leading to the development of a wide range of code benchmarks.
Existing benchmarks primarily focus on code generation and program repair tasks at different granularities~\cite{SDLC_Survey_Benechmarks_CLLMs_and_Agent}.
Function-level benchmarks~\cite{SDLC_Survey_Benechmarks_CLLMs_and_Agent} such as HumanEval~\cite{chen2021evaluating} and MBPP~\cite{MBPP} focus on generating correct implementations for short and self-contained programming problems.
Class-level benchmarks~\cite{SDLC_Survey_Benechmarks_CLLMs_and_Agent}, including ClassEval~\cite{classeval}, evaluate object-oriented code generation involving class design and method implementation.
Repository-level benchmarks~\cite{SDLC_Survey_Benechmarks_CLLMs_and_Agent} such as HumanEvo~\cite{humanevo} and ConCode~\cite{concode} aim to assess code generation in more realistic development settings with broader context.

Despite their success in evaluating functional correctness and general programming ability, these benchmarks are largely correctness-oriented and are not designed to evaluate performance optimization on computationally intensive HPC kernels, where efficiency is tightly coupled with hardware characteristics, memory behavior, and parallel execution.

\paragraph{LLMs for Code Optimization and Kernel-Level Benchmarks.}

Recent efforts have tapped into the use of LLMs for optimizing code performance, ranging from kernel-level to application-level optimization~\cite{Challenges_and_path_towards_AI}.
Prior studies such as PIE~\cite{PIE} and AlphaDev~\cite{alphadev} demonstrate that LLMs can discover efficient implementations for specific algorithms or low-level routines.
More recently, kernel-centric benchmarks such as KernelBench~\cite{kernelbench} and TritonBench~\cite{tritonbench} have been proposed to evaluate LLMs on generating high-performance GPU kernels, particularly for machine learning workloads.
ParEval~\cite{pareval} evaluates LLM-generated parallel code across scientific computing tasks and programming models, reporting both correctness and performance metrics, including $\mathrm{@k}$ and $\mathrm{speedup}_n@k$.
ENAMEL~\cite{enamel} evaluates generated code efficiency through $\mathrm{eff}@k$ and expert-optimized reference implementations.
Unlike KernelBench~\cite{kernelbench} and TritonBench~\cite{tritonbench}, which evaluate standalone GPU kernels primarily for ML workloads, HPC-Bench focuses on profiling-identified main computational kernels from real HPC benchmark programs.
ParEval~\cite{pareval} and ENAMEL~\cite{enamel} instead evaluate newly generated solutions rather than local optimizations of existing HPC code.

\paragraph{HPC Performance Benchmarking.}

The HPC community has long relied on benchmark suites to evaluate architectures, compilers, and runtime systems under controlled and reproducible workloads.
Representative suites span both kernel-centric and application-level settings, including PolyBench~\cite{polybench} for static-control-flow numerical kernels, Rodinia~\cite{rodinia} for CPU/GPU workloads, PARSEC~\cite{parsec} for multithreaded applications, MiBench~\cite{mibench} for embedded programs, and accelerator-centric suites such as MachSuite~\cite{machsuite} and Parboil~\cite{parboil} for CUDA/OpenCL implementations for CPU/GPU performance comparison.
Recent LLM-based HPC systems further highlight the importance of measurement-in-the-loop evaluation: ParaCodex~\cite{paracodex}, for example, operationalizes an HPC-engineering workflow with compilation/testing gates and profiling-driven feedback, using profiler-reported GPU time (kernel execution plus transfers) as a performance signal;
AlphaEvolve~\cite{alphaevolve} also explores iterative, evaluator-guided code improvements for scientific and computational problems.
Such systems are complementary to our setting: they study agentic, feedback-driven optimization workflows, whereas HPC-Bench isolates the single-model local optimization setting under a fixed sampling budget.
In contrast to traditional benchmarking or autotuning, HPC-Bench treats LLMs as stochastic local optimizers operating on annotated \emph{main computational kernels} and evaluates the resulting replacements within the original benchmark harness using automated program-level correctness verification and kernel-runtime metrics.

\section{\HPCBench: A Framework for HPC Code Optimization}
\label{sec:hpcbench}

\HPCBench evaluates LLMs as local optimizers of profiling-identified computational kernels embedded in complete HPC programs.
The suite contains $92$ CPU/OpenMP workloads across $16$ computational motifs and $22$ benchmarks for CUDA.
It standardizes kernel annotations, inputs, builds, correctness checking, and correctness-gated performance metrics.

\subsection{Overview and Pipeline}
\label{overview}

\begin{figure*}[!htbp]
    \centering
    \includegraphics[width=\linewidth]{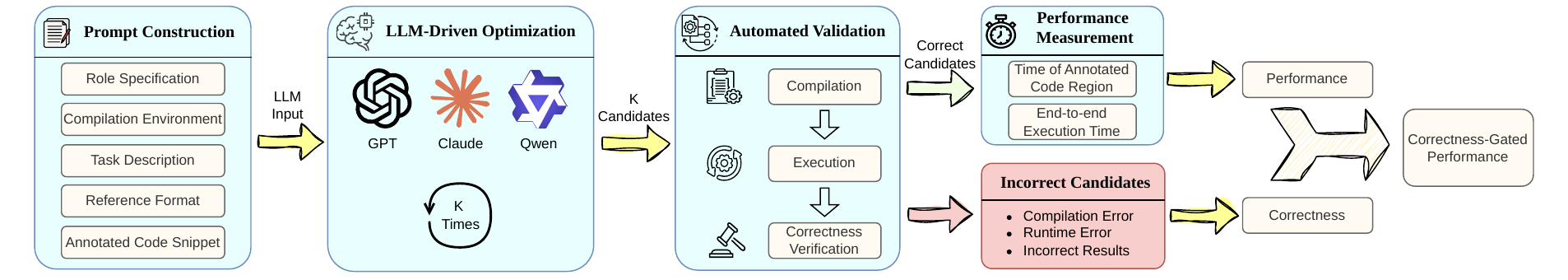}
    \caption{HPC-Bench Workflow.}
    \label{fig:overview}
\end{figure*}

Given a benchmark with curated optimization metadata, \HPCBench constructs a benchmark-specific prompt containing the LLM role, compilation environment, optimization setting, task constraints, required output format, annotated code region, and minimal compilation context.
For each model and benchmark, \HPCBench samples $N$ candidate replacements from the same prompt.
Each candidate is extracted, spliced back into the original benchmark source, compiled, executed, checked against the baseline output, and timed, as illustrated in~\Cref{fig:overview}.
Candidates that fail compilation, execution, or correctness checking are recorded as failures; only correct candidates contribute to the correctness-gated speedup metrics in~\Cref{subsec:evaluation-metrics}.

\subsection{Task Format} 
\label{task-format}





\begin{lstlisting}[
  style=myscreenshot_small,
  language={},
  float=tbp,
  caption={Input Format},
  captionpos=b,
  label={lst:task_format},
  aboveskip=6pt,
  belowskip=6pt,
  xleftmargin=0.5em
]
[System Prompt]
You are a code optimization assistant.
Output compilable code only.
(e.g., target hardware, compiler, and OS specified.)
[Task Prompt]
Optimize the code below under given constraints
(e.g., serial/OpenMP/CUDA).
[Input Code]
/* START_LINE */
void atmux(...) {
  // main computational kernel
}
/* END_LINE */
[Output Format]
<<<CODE>>>
<optimized code between START_LINE and END_LINE>
<<<END-CODE>>>
\end{lstlisting}

\HPCBench poses kernel-centric optimization tasks.
The LLM rewrites only the annotated editable region while the original benchmark scaffold remains responsible for input construction, data-structure initialization, compilation, execution, and program-level correctness checking.
The editable region contains at least one complete function; in a few benchmarks, it spans multiple closely related functions when the performance-critical logic cannot be optimized in isolation.
The generated replacement is wrapped in fixed delimiters for automatic extraction and insertion, as illustrated in~\Cref{lst:task_format}.
The full prompt templates are provided in~\Cref{app_sec:prompt_template}.

\subsection{Benchmark Suite}
\label{benchmark-suite}

\paragraph{Benchmark Sources and Coverage.}
The benchmarks in \HPCBench are curated from widely used HPC benchmark suites, including Rodinia~\cite{rodinia}, PolyBench~\cite{polybench}, MiBench~\cite{mibench}, MachSuite~\cite{machsuite}, PARSEC~\cite{parsec}, Codee~\cite{codee_benchmark}, and Parboil~\cite{parboil}.
They cover $16$ computational motifs: Dense Linear Algebra (DLA), Sparse Linear Algebra (SLA), Stencil/Scientific Computing (SC), Graph Algorithms (GA), Dynamic Programming (DP), N-body Methods (NBM), Image and Video Processing (IVP), Clustering/Machine Learning (CML), Statistical Computations (STC), Spectral Methods (SM), Computational Fluid Dynamics (CFD), Monte Carlo Methods (MCM), Cryptography/Encoding (CE), Sorting/Search (SS), Medical/Scientific Computing (MSC), and Utility Benchmarks (UB).

The selection is intended to cover representative HPC workloads rather than randomly sample a single suite.
Within each motif, benchmarks differ in access pattern, dependency structure, and optimization opportunity: 
DLA includes dense matrix kernels such as \texttt{gemm}, \texttt{sgemm}, and \texttt{matmul}, where blocking and cache reuse are central, dense matrix-vector kernels such as \texttt{atax}, \texttt{bicg}, and \texttt{gesummv}, but also factorization and triangular kernels such as \texttt{cholesky}, \texttt{lu}, \texttt{ludcmp}, and \texttt{trisolv}, where dependencies constrain valid transformations;
SLA consists of format-specific SpMV kernels (\texttt{spmv}, \texttt{spmv-crs}, and \texttt{spmv-ellpack}), where indirect indexing limits simple vectorization;
SC spans 1D to 3D structured-grid kernels such as \texttt{jacobi-1d}, \texttt{jacobi-2d}, \texttt{hotspot}, and \texttt{heat-3d}, which differ in dimensionality, neighbor structure, and update dependencies.
Extended motif-level details are reported in~\Cref{app_sec:motif_characterization}.

For CUDA-based optimization, we construct a dedicated CUDA benchmark set drawn from Rodinia~\cite{rodinia}.
This set partially overlaps with the Rodinia benchmarks used in the CPU/OpenMP evaluations, while additionally including five GPU-oriented workloads (\texttt{dwt2d}, \texttt{nn}, \texttt{gaussian}, \texttt{huffman}, and \texttt{hybridsort}) that are not included in the CPU/OpenMP version.

\paragraph{Kernel Selection and Context Curation.}
We seed kernel selection with external profilers: HPCToolkit~\cite{HPCTOOLKIT} for CPU/OpenMP workloads and Nsight Systems~\cite{nsight-systems} for CUDA workloads.
We then manually filter non-computational hotspots such as printing, file I/O, initialization, timing instrumentation, and cleanup code.
The optimization target is usually the hottest computational function after filtering; when the hotspot spans tightly coupled caller-callee logic, such as in \texttt{b+tree}, we annotate the smallest necessary multi-function region.

For each selected target, we curate the minimal compilation context required by the annotated region, including helper functions, macros/constants, typedefs/structs, and headers.
This context is included in the prompt, while the generated replacement is inserted back into the unmodified benchmark source.
Thus, the LLM optimizes a local code region while the full benchmark driver provides the execution context and correctness verification.

\paragraph{Structural Complexity Stratification.}
Each benchmark is assigned to a structural complexity stratum from \textbf{d1} (lowest structural complexity) to \textbf{d4} (highest structural complexity) using a static structural complexity proxy.
Let $b$, $\ell$, $d$, and $f$ denote branch count, lines of code (LOC), maximum loop nesting depth, and function count within the annotated region, respectively.
The branch count includes \texttt{if}, \texttt{case}, and ternary operators.
Each raw metric $x$ is min-max scaled to $[0,10]$ using the $5$th and $95$th percentiles across the $92$ CPU/OpenMP benchmarks:
\begin{equation}
\hat{x} =
\min\!\left(10,\,
\max\!\left(0,\,
\frac{x-x_5}{x_{95}-x_5}\cdot 10
\right)\right).
\end{equation}
The composite score is
\begin{equation*}
\mathrm{score}=0.45\hat{b}+0.25\hat{\ell}+0.20\hat{d}+0.10\hat{f}.
\end{equation*}
We bucket the score as $d_1=[0,2.5)$, $d_2=[2.5,5)$, $d_3=[5,7.5)$, and $d_4=[7.5,10]$.
Loop nesting depth and function count are extracted using Lizard; LOC and branch count are computed by custom parsers on the annotated region.
This score is a static proxy for optimization structural complexity:
it correlates with optimization outcomes but is not a direct measure of headroom, which also depends on baseline runtime, memory behavior, and available parallelism.
\Cref{tab:difficulty-statistics} reports the per-category statistics for the $92$ CPU/OpenMP benchmarks.
The CUDA benchmark labels in \Cref{app_sec:benchmark_list} are assigned using the same structural complexity definition for metadata consistency, but the CUDA set is not used for structural complexity stratum analysis because it contains only $22$ benchmarks.

\begin{table}[tp]
\centering
\small
\renewcommand{\arraystretch}{1.15}
\caption{Structural Complexity Stratum Statistics}
\label{tab:difficulty-statistics}
\begin{tabular*}{\linewidth}{@{\extracolsep{\fill}}lcccc}
\hline
\textbf{Structural Complexity} & \textbf{\#LOC} & \textbf{\#func} & \textbf{\#branch} & \textbf{\#loop} \\ \hline
$d_1$ (66.3\%) & 34.34 & 1.46 & 0.72 & 2.61 \\
$d_2$ (18.5\%) & 100.47 & 1.71 & 5.18 & 3.12 \\
$d_3$ (9.8\%)  & 139.33 & 2.22 & 11.89 & 4.56 \\
$d_4$ (5.4\%)  & 316.00 & 4.00 & 19.80 & 19.00 \\ \hline
\end{tabular*}
\end{table}

\paragraph{Standardization across Benchmarks.}
To reduce confounders, we remove pre-existing OpenMP parallelization to enforce a uniform serial baseline for CPU experiments.
We standardize inputs into five scales: \emph{mini}, \emph{small}, \emph{medium}, \emph{large}, and \emph{extra-large}; when original benchmark inputs are unavailable, we construct missing splits manually and validate that the baseline program produces valid reference outputs.
LLMs are not used for input generation or correctness judging.
Runtime is measured with warmup runs followed by repeated timed trials, using the annotated kernel mean runtime as the primary signal.
Unless otherwise stated, the main paper reports metrics using annotated kernel runtime; total runtime is recorded for \Cref{app_subsec:total_runtime_metrics} reporting.

\subsection{Evaluation Metrics} 
\label{subsec:evaluation-metrics}

\HPCBench evaluates models with correctness-gated, sampling-aware metrics that separate \emph{reliability} from the \emph{magnitude} of achieved speedup.
Intuitively, $\mathrm{Fast}@k$ measures the probability that at least one candidate among $k$ samples is both correct and sufficiently fast, while $\mathrm{Speedup}@k$ measures the expected best correctness-gated speedup among $k$ samples.
Full notation, auxiliary statistics, and extended visualization details are deferred to~\Cref{app_sec:metric_full}.

For benchmark $b$, candidate $j$, the number of threads $n$, and runtime type $t\in\{\mathrm{kernel},\mathrm{total}\}$, we define correctness-gated speedup as
\begin{equation}
S^{(t)}_{b,j,n}
=
\left(\hat{T}_{b,n}^{(t)} \big/ T^{(t)}_{b,j,n}\right)
\cdot
\mathbf{1}\{\mathrm{correct}_{b,j}=1\},
\end{equation}
where $\hat{T}_{b,n}^{(t)}$ is the baseline runtime, $T^{(t)}_{b,j,n}$ is the runtime of candidate $j$, and incorrect implementations receive zero speedup.

\paragraph{$\mathrm{Fast}@k$.}
A threshold $p_{\mathrm{thr}}=1.0$ requires a correct non-slowdown, while $p_{\mathrm{thr}}>1.0$ imposes a minimum speedup requirement.
Given a speedup threshold $p_{\mathrm{thr}}$, let
\begin{equation}
c^{(t)}_{b,n}(p_{\mathrm{thr}})
=
\sum_{j=1}^{N}
\mathbf{1}\!\left\{
S^{(t)}_{b,j,n}\ge p_{\mathrm{thr}}
\right\}
\end{equation}
be the number of generated candidates that are both correct and achieve a correctness-gated speedup of at least $p_{\mathrm{thr}}$.
Sampling $k\le N$ candidates uniformly without replacement,
\begin{equation}
\mathrm{Fast}^{(t)}_{n,p_{\mathrm{thr}}}@k(b)
=
\Pr\!\left(
\max_{j\in\mathcal{K}} S^{(t)}_{b,j,n}
\ge p_{\mathrm{thr}}
\right)
=
1-
\binom{N-c^{(t)}_{b,n}(p_{\mathrm{thr}})}{k}
\Big/
\binom{N}{k}.
\end{equation}

\paragraph{$\mathrm{Speedup}@k$.}
While $\mathrm{Fast}@k$ captures the probability of finding a qualifying candidate, it does not capture the magnitude of the best speedup.
We therefore define
\begin{equation}
\mathrm{Speedup}^{(t)}_{n,p_{\mathrm{thr}}}@k(b)
=
\mathbb{E}\!\left[
\max_{j\in\mathcal{K}} S^{(t)}_{b,j,n}
\cdot
\mathbf{1}
\left\{
\max_{j\in\mathcal{K}} S^{(t)}_{b,j,n}
\ge p_{\mathrm{thr}}
\right\}
\right].
\end{equation}
A benchmark whose sampled candidates do not satisfy $p_{\mathrm{thr}}$ contributes $0$, not the baseline value $1.0$.
This convention makes $\mathrm{Speedup}@k$ monotonically non-increasing as $p_{\mathrm{thr}}$ increases.
We compute this expectation exactly by enumerating the $\binom{N}{k}$ subsets.

Unless otherwise stated, reported aggregate metrics are macro-averaged across benchmarks within the corresponding stratum, such as motif, structural complexity stratum, or dataset size, when applicable.
\begin{figure*}[t]
  \centering
  \includegraphics[width=\textwidth]{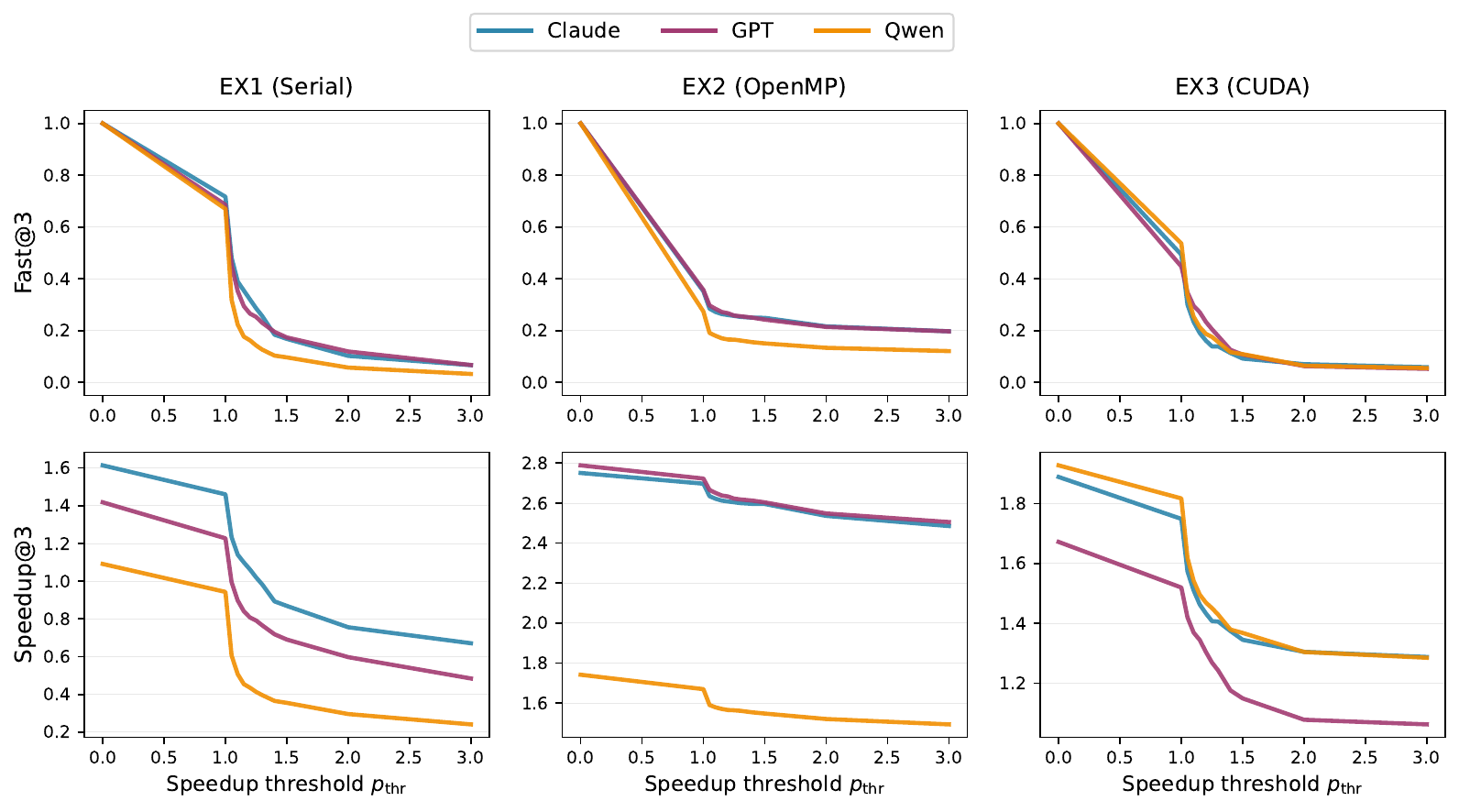}
  \caption{$\mathrm{Fast}@3$/$\mathrm{Speedup}@3$ versus the speedup threshold $p_{\mathrm{thr}}$, aggregated over all benchmarks and dataset sizes for EX1–EX3.}
  \label{fig:rq1_rq2}
\end{figure*}

\section{Experimental Setup}
\label{sec:experimental_setup}

We conduct experiments on \HPCBench to evaluate the ability of LLMs to optimize performance-critical code under realistic workflows.
Specifically, we evaluate three settings that differ in their optimization constraints and target hardware: serial CPU optimization (EX1), OpenMP CPU parallelization (EX2), and CUDA GPU optimization (EX3).

\paragraph{Optimization Settings.}
EX1 permits only single-thread CPU transformations, such as loop restructuring, scalar hoisting, and memory-access optimization.
EX2 extends EX1 with OpenMP parallelization, while forbidding in-source thread-count control; instead, thread counts ($\{1,2,8,16,32\}$) are set externally via \texttt{OMP\_NUM\_THREADS}.
EX1 and EX2 use the same 92-benchmark suite.
EX3 evaluates CUDA optimization on the 22-benchmark CUDA set drawn from Rodinia.

\paragraph{Models and Sampling.}
We evaluate GPT-5.1~\cite{openai_gpt51}, Claude Sonnet~4.5~\cite{anthropic_claude45}, and Qwen3-Coder-480B-A35B-Instruct~\cite{qwen3_coder_480b}.
A per-setting API cost breakdown is provided in~\Cref{app_sec:budget}.
For each benchmark, setting, and model, we generate $N=10$ candidates using the same prompt and decoding protocol.
Unless otherwise stated, we report sampling-aware metrics at $k=3$ across speedup thresholds $p_{\mathrm{thr}}$.
For EX2, we report the 16-thread setting in the main paper and defer the full $\{1,2,8,16,32\}$ sweep to~\Cref{app_subsec:thread_sensitivity}.

\paragraph{Hardware.}
EX1 and EX2 run on a dual-socket AMD EPYC 7543 system with 64 physical cores and simultaneous multithreading disabled, compiled with GCC/G++ 14.2 on Enterprise Linux 8.
EX3 runs on an NVIDIA A100 GPU, with CUDA code compiled by \texttt{nvcc} 12.4.
Host code is compiled with the same compiler toolchain used in the CPU experiments.
Full per-benchmark compilation and link flags are detailed in~\Cref{app_subsec:compilation_flags}.

\section{Evaluation}
\label{sec:evaluation}

We evaluate LLMs on \HPCBench from three complementary perspectives: reliability, magnitude of performance gains, and robustness across motifs, structural complexity strata, and sampling budgets.

\subsection{Research Questions}

Our evaluation is guided by three questions:
\begin{itemize}
    \item \textbf{RQ1 (Reliability).} How reliably can models produce \emph{correct and sufficiently fast} optimizations under a limited sampling budget?
    \item \textbf{RQ2 (Performance Gains).} When useful optimizations are found, how large are the achieved performance gains?
    \item \textbf{RQ3 (Sensitivity and Robustness).} How do reliability and performance gains vary across computational motifs and structural complexity strata, and as the sampling budget changes?
\end{itemize}

\subsection{RQ1: Reliability Under a Sampling Budget}
\label{subsec:rq1_reliability}

\textbf{Answer: Models can often find correct or modestly faster candidates, but $\mathrm{Fast}@3$ drops sharply as the required speedup threshold increases, especially in the CUDA setting.}

The top row of~\Cref{fig:rq1_rq2} reports $\mathrm{Fast}@3$ across speedup threshold $p_{\mathrm{thr}}$ for EX1-EX3, macro-averaged over all benchmarks and the five standardized dataset sizes.
Across all models and execution settings, $\mathrm{Fast}@3$ decreases sharply as the required $p_{\mathrm{thr}}$ increases.
At $p_{\mathrm{thr}}=1.0$, serial CPU optimization remains relatively reliable: averaged across the five dataset sizes, $\mathrm{Fast}@3$ is $0.72$ for Claude Sonnet~4.5, $0.69$ for GPT-5.1, and $0.67$ for Qwen.
However, requiring at least $1.5{\times}$ speedup reduces EX1 reliability to $0.17$, $0.17$, and $0.10$, respectively.

The same threshold sensitivity appears in EX2 and EX3.
For EX2 at 16 threads, average $\mathrm{Fast}@3$ drops from $0.35$-$0.36$ for Claude Sonnet~4.5/GPT-5.1 at $p_{\mathrm{thr}}=1.0$ to roughly $0.24$-$0.25$ at $p_{\mathrm{thr}}=1.5$, while Qwen3-Coder decreases from $0.27$ to $0.15$.
EX3 is especially threshold-sensitive: although average $\mathrm{Fast}@3$ at $p_{\mathrm{thr}}=1.0$ is still $0.45$-$0.54$, it falls to about $0.09$-$0.11$ at $p_{\mathrm{thr}}=1.5$; across all model-dataset pairs, the range at this threshold is only $0.05$--$0.17$.
At high thresholds, the model curves also become close to each other, suggesting that CUDA optimization is dominated by the structural complexity of producing any correct and sufficiently fast candidate rather than by small differences in model ranking.


\subsection{RQ2: Correctness-Gated Performance Gains}
\label{subsec:rq2_speedup}

\textbf{Answer: \HPCBench contains genuine high-speedup opportunities, but correctness-gated gains are concentrated in a subset of workloads that current LLMs can optimize reliably.}

The bottom row of~\Cref{fig:rq1_rq2} reports $\mathrm{Speedup}@3$ as a function of $p_{\mathrm{thr}}$.
At $p_{\mathrm{thr}}=1.0$, EX1 achieves average correctness-gated speedups of $1.46{\times}$ for Claude Sonnet~4.5, $1.23{\times}$ for GPT-5.1, and $0.94{\times}$ for Qwen3-Coder.
At $p_{\mathrm{thr}}=1.5$, these values decrease to $0.87{\times}$, $0.69{\times}$, and $0.36{\times}$, respectively, because many benchmarks no longer contribute a qualifying candidate.
EX3 follows a similar threshold-sensitive pattern: at $p_{\mathrm{thr}}=1.0$, average $\mathrm{Speedup}@3$ ranges from $1.52{\times}$ to $1.82{\times}$, but at $p_{\mathrm{thr}}=1.5$ it drops to $1.15$--$1.37{\times}$.

EX2 behaves differently.
At $p_{\mathrm{thr}}=1.0$, Claude and GPT-5.1 reach $2.70{\times}$ and $2.72{\times}$ on average, while Qwen3-Coder reaches $1.67{\times}$; at $p_{\mathrm{thr}}=1.5$, these values remain comparatively stable at $2.60{\times}$, $2.60{\times}$, and $1.55{\times}$.
This slower decay does not mean that EX2 is uniformly easy.
Rather, the EX2 \emph{Best-of-$N$} kernel speedup distribution (\Cref{app_subsec:ex2_distribution}) is highly polarized: a small set of highly parallelizable kernels achieve very large speedups and survive all thresholds, while many short-running kernels fail to exceed even modest thresholds because OpenMP overhead dominates execution.
As a result, $\mathrm{Fast}@3$ drops with increasing $p_{\mathrm{thr}}$, but $\mathrm{Speedup}@3$ remains relatively flat because the macro-average continues to be anchored by the same high-contributing kernels, e.g., \texttt{matmul} ($49.6{\times}$), \texttt{haccmk} ($30.7{\times}$), \texttt{coulomb} ($13.7{\times}$), \texttt{hotspot3D} ($13.3{\times}$), and \texttt{particlefilter} ($12.4{\times}$).

\paragraph{Representative High-speedup Cases.}
The aggregate curves are pulled up by a small set of benchmarks where models discover qualitatively stronger transformations rather than incremental cleanups.
We highlight three cases from Claude Sonnet~4.5.
In EX1, \texttt{matmul} reaches $4.3{\times}$ speedup by applying $64{\times}64$ loop tiling, changing the loop order from \texttt{i-j-k} to \texttt{i-k-j}, and hoisting \texttt{A[i][k]} into a scalar, improving cache locality and removing redundant loads.
In EX2, \texttt{hotspot3D} reaches $13.3{\times}$ speedup by adding \texttt{\#pragma omp parallel for collapse(2) schedule(static)}, hoisting time-step-invariant expressions such as \texttt{dt / Cap} and \texttt{ct * amb\_temp}, and precomputing neighbor offsets for the innermost stencil loop.
In EX3, \texttt{particlefilter} reaches $21.9{\times}$ speedup by replacing a per-thread linear CDF scan with binary search, reducing the search complexity from $O(N)$ to $O(\log N)$.

These cases illustrate three recurring sources of large gains in \HPCBench: cache/locality-aware loop restructuring, OpenMP parallelization with overhead reduction, and improvements in algorithmic complexity.

\subsection{RQ3: Robustness Across Motifs, Structural Complexity, and Sampling Budgets}
\label{subsec:motif_difficutly_level}

\textbf{Answer: Reliability and speedup vary across computational motifs, structural complexity strata, and sampling budgets. Compute-heavy and data-parallel motifs benefit most, while irregular access patterns, dependencies, and high structural complexity code reduce optimization success; increasing the sampling budget mainly helps models find stronger candidates, but only modestly changes whether any qualifying candidate exists.}

\subsubsection{Motif-level Outcomes Track Motif-level Structure}
\label{subsubsec:motif-level_sensitivity}

Motif-level results show that the aggregate trends are not uniform across the suite.
On the medium dataset at $p_{\mathrm{thr}}=1.0$, averaged across the three models, EX1 achieves the highest $\mathrm{Speedup}@3$ on NBM ($2.56{\times}$), SC ($2.46{\times}$), MCM ($1.51{\times}$), and DLA ($1.35{\times}$).
These motifs expose transformations such as loop reordering, scalar hoisting, locality improvement, and arithmetic restructuring.
By contrast, motifs such as CFD ($0.05{\times}$), IVP ($0.46{\times}$), MSC ($0.59{\times}$), SM ($0.61{\times}$), SLA ($0.65{\times}$), GA ($0.69{\times}$), and CML ($0.70{\times}$) show more limited gains, often because the bottleneck is dominated by memory traffic, irregular access, or benchmark-specific data dependencies.

In EX2, the top motifs at 16 threads are NBM ($7.17{\times}$), DLA ($2.30{\times}$), MCM ($2.23{\times}$), IVP ($1.74{\times}$), and CFD ($1.65{\times}$).
This confirms that motifs with abundant data parallelism and sufficient runtime scale benefit most from OpenMP.
In contrast, SLA and SM both average near zero, while SS ($0.16{\times}$), STC ($0.16{\times}$), CE ($0.21{\times}$), and UB ($0.77{\times}$) are muted by limited parallelism, irregular access, or overhead-sensitive execution.
Selected motif-level visualizations for $\mathrm{Fast}@3$ and $\mathrm{Speedup}@3$ are reported in~\Cref{app_subsec:stratified_result}; EX3 motif-level macro-averages are omitted from the main analysis because the 22-benchmark CUDA benchmark set is too small for stable per-motif conclusions.

\paragraph{What Models Try, and Where They Fail, per Motif.}
Generated programs show motif-conditioned behavior.
In DLA and SC, models often attempt loop reordering, scalar extraction, pointer caching, and constant hoisting, but failures arise when factorization dependencies or floating-point-order changes break correctness.
In SLA and GA, models are more conservative: indirect sparse layouts, pointer chasing, and traversal dependencies make deeper restructuring risky.
This explains why the same model performs better on compute-regular kernels than on irregular or dependency-heavy workloads.

\subsubsection{Structural Complexity Stratum Analysis: Structural Complexity Interacts with Optimization Headroom}
\label{subsubsec:difficulty-level_sensitivity}

For EX1 and EX2, the structural complexity score predicts optimization outcomes in some settings, but its relationship with performance is also influenced by runtime-scale factors. 
We omit EX3 because its $22$-benchmark CUDA set is too small for stable structural complexity stratum analysis.
In EX1, strata with lower structural complexity generally have higher $\mathrm{Fast}@3$ and $\mathrm{Speedup}@3$ than strata with higher structural complexity, suggesting that simpler annotated regions are easier for LLMs to optimize correctly. 
However, EX2 demonstrates that structural simplicity alone does not imply large parallel optimization headroom.
At 16 threads on the medium dataset, $41$ out of $61$ $d_1$ benchmarks ($67\%$) fail to reach even $1.05{\times}$ speedup, even though $d_1$ is structurally simple.
Their median baseline kernel runtime is only $7.33$\,ms, and several representative cases are much shorter: \texttt{gesummv} runs in roughly $152\,\mu$s, \texttt{stencil-2d-mach} in roughly $11\,\mu$s, and \texttt{spmv-crs-mach} in roughly $3.8\,\mu$s.
For such kernels, OpenMP overhead dominates computation, so parallelization can easily slow the program down.

By contrast, $d_2$ and $d_3$ benchmarks have longer median baseline kernel runtimes ($1.05$\,s and $1.37$\,s, respectively) and therefore provide more opportunity for thread-level parallelism.
Only $5/17$ $d_2$ benchmarks and $1/9$ $d_3$ benchmarks fail to reach $1.05{\times}$ speedup, and representative kernels such as \texttt{hotspot} ($5.34{\times}$), \texttt{b+tree} ($8.49{\times}$), \texttt{nw} ($7.39{\times}$), and \texttt{hotspot3D} ($13.27{\times}$) show substantial OpenMP gains.
The $d_4$ group is small ($5$ benchmarks) but exhibits heterogeneous outcomes that reinforce the score's role as a complexity proxy.
Longer baseline runtimes coexist with correctness-side failures: \texttt{cutcp} and \texttt{lbm} produce no correct candidates, while \texttt{heartwall}, \texttt{susan-c-mibench}, and \texttt{susan-e-mibench} reach substantial speedups of up to $8.79{\times}$, $7.33{\times}$, and $8.34{\times}$, respectively.
Hence $d_1$ and the hardest $d_4$ cases fail for different reasons: 
$d_1$ is often overhead-limited by short kernel runtimes, whereas the hardest 
$d_4$ cases are more often correctness-limited by structural complexity.
An aggregate breakdown by structural complexity stratum supporting this interpretation is provided in~\Cref{app_subsec:stratified_result}, with a focused diagnostic of the $d_4$ benchmarks in~\Cref{tab:d4_breakdown}.

\subsubsection{Sampling-Budget Sensitivity}
\label{subsubsec:sampling_budget_sensitivity}


\begin{figure}[!htbp]
  \centering
  \captionsetup{justification=centering}
  \captionsetup{justification=centering}
  \begin{subfigure}{0.48\columnwidth}
    \centering
    \includegraphics[width=\columnwidth]{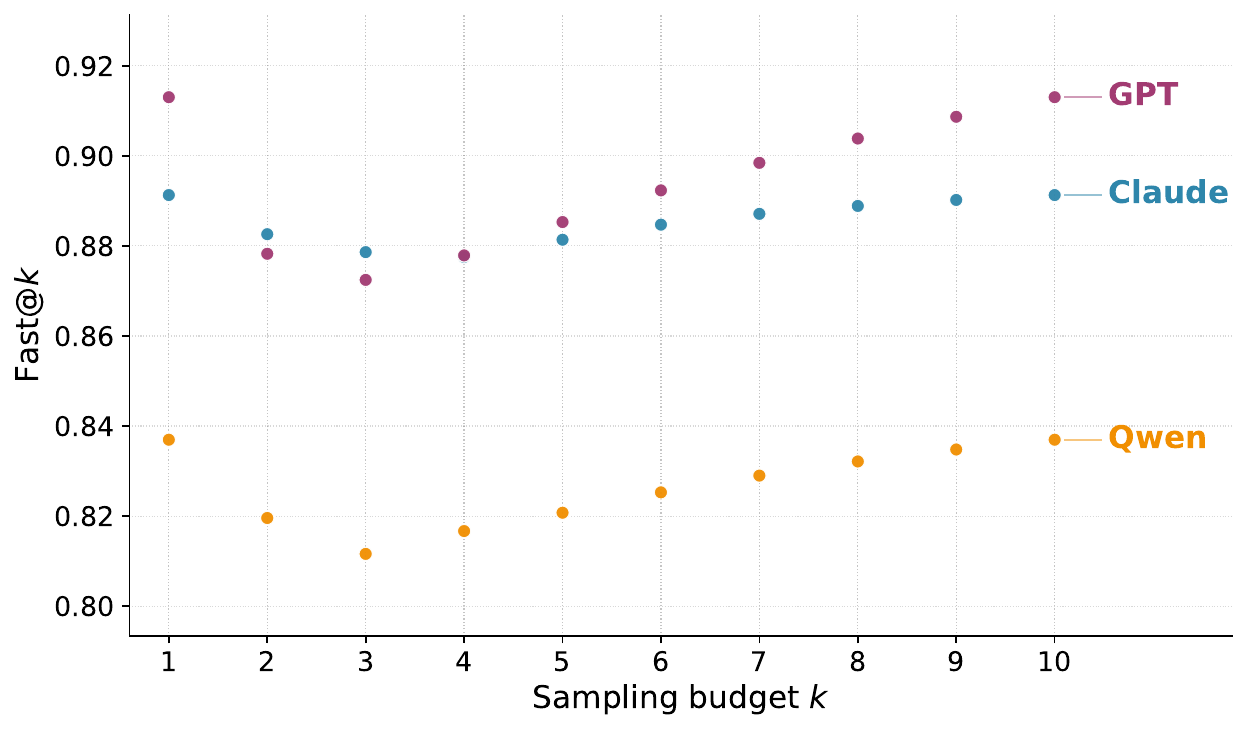}
    \caption{$\mathrm{Fast}@k$ versus $k$}
    \label{fig:sensitivity_ex1_fast}
  \end{subfigure}
  \hspace{0.02\columnwidth}
  \begin{subfigure}{0.48\columnwidth}
    \centering
    \includegraphics[width=\columnwidth]{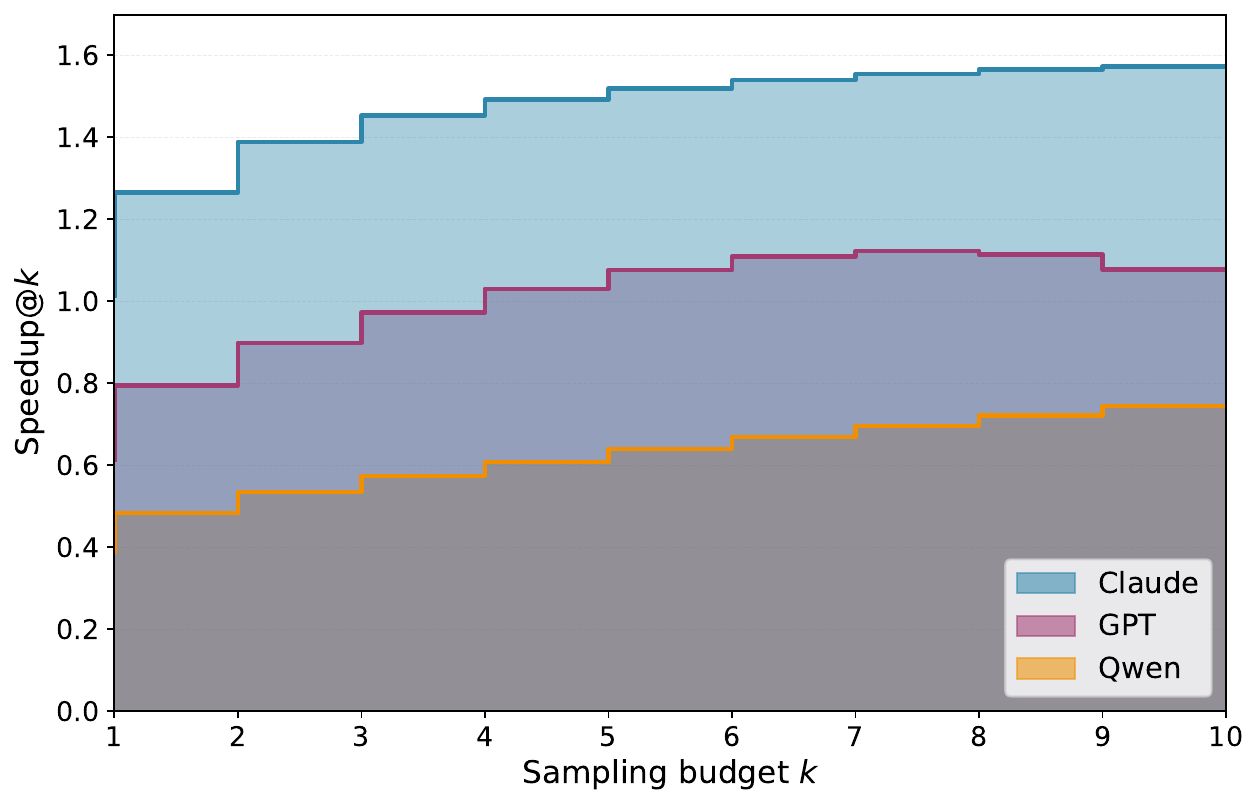}
    \caption{$\mathrm{Speedup}@k$ versus $k$}
    \label{fig:sensitivity_ex1_speedup}
  \end{subfigure}
  \caption{$\mathrm{Fast}@k$ and $\mathrm{Speedup}@k$ versus $k$ for EX1 when $p_{\mathrm{thr}}=1.05$.}
  \label{fig:sensitivity_ex1}
\end{figure}

\Cref{fig:sensitivity_ex1} reports $\mathrm{Fast}@k$ and $\mathrm{Speedup}@k$ as functions of the sampling budget $k$ for EX1 on the medium dataset at $p_{\mathrm{thr}}=1.05$.
Increasing $k$ has only a modest effect on $\mathrm{Fast}@k$: all three models remain in a relatively narrow band, indicating that most benchmarks that can produce at least one correct and slightly faster candidate already do so with a small number of samples.
In contrast, $\mathrm{Speedup}@k$ grows more noticeably with $k$, especially for Claude Sonnet~4.5, showing that additional samples are more useful for finding a stronger candidate than for substantially changing the probability of finding any qualifying candidate.
Corresponding sensitivity curves for EX2 and EX3 are reported in~\Cref{app_subsec:sampling_budget_sensitivity}.

\subsection{Correctness and Failure Modes}
\label{subsec:correctness_failure_modes}

\begin{table}[H]
\centering
\captionsetup{justification=centering}
\caption{Correctness and failure statistics across optimization settings.
CF denotes compilation failures, RF denotes runtime failures, and CE denotes incorrect executions.
Percentages are computed over all generated trials within each EX setting.}
\label{tab:correctness_failure}
\setlength{\tabcolsep}{6pt}
\small
\begin{tabular}{
    l
    S[table-format=2]
    S[table-format=4]
    S[table-format=3]
    S[table-format=2.1]
    S[table-format=2]
    S[table-format=1.1]
    S[table-format=3]
    S[table-format=2.1]
    S[table-format=4]
    S[table-format=2.1]
}
\toprule
\textbf{EX} &
\textbf{\#Bench.} &
\textbf{Trials} &
\textbf{CF} &
\textbf{CF (\%)} &
\textbf{RF} &
\textbf{RF (\%)} &
\textbf{CE} &
\textbf{CE (\%)} &
\textbf{Correct} &
\textbf{Correct (\%)} \\
\midrule
EX1 & 92 & 2760 & 249 & 9.0 & 50 & 1.8 & 339 & 12.3 & 2122 & 76.9 \\
EX2 & 92 & 2760 & 574 & 20.8 & 13 & 0.5 & 477 & 17.3 & 1696 & 61.4 \\
EX3 & 22 & 660  & 232 & 35.2 & 32 & 4.8 & 140 & 21.2 & 256  & 38.8 \\
\bottomrule
\end{tabular}
\end{table}

\Cref{tab:correctness_failure} summarizes correctness outcomes across EX1--EX3, with detailed failure breakdowns in~\Cref{app_subsec:failure_breakdown}.
Correctness decreases from $76.9\%$ in EX1 to $61.4\%$ in EX2, and $38.8\%$ in EX3, reflecting the additional challenges of preserving semantics under OpenMP parallelism and CUDA-specific memory/synchronization constraints.

Most failures are not simple formatting artifacts.
Compilation failures are dominated by syntax/type/declaration errors, while malformed outputs account for $8.0\%$, $3.5\%$, and $34.1\%$ of compilation failures in EX1--EX3, respectively.
All runtime failures in EX1 and EX2 are memory errors, whereas $96.9\%$ of runtime failures in EX3 are CUDA errors.
Correctness errors include floating-point mismatches, element mismatches, and missing outputs; 
incorrect outputs account for $79.1\%$, $86.4\%$, and $100.0\%$ of correctness errors in EX1--EX3, respectively.
Thus, simple output post-processing can address only a limited subset of failures.

\section{Conclusions}

We present \HPCBench, an end-to-end benchmark for evaluating LLMs as performance optimizers of profiling-identified computational kernels embedded in realistic HPC programs. 
\HPCBench combines standardized benchmark construction, automated compilation and correctness checking, reproducible runtime measurement, and correctness-gated metrics that jointly capture optimization reliability and achieved speedup under a fixed sampling budget.
Across serial CPU, OpenMP, and CUDA settings, our evaluation shows that current LLMs can occasionally discover substantial performance improvements, including locality-aware loop restructuring, effective OpenMP parallelization, and algorithmic improvements. 
However, these gains are unevenly distributed and are constrained by reliability: 
stricter speedup thresholds, structural complexity, and more complex optimization settings make it harder to obtain a correct and sufficiently fast candidate.
Motif level and structural complexity stratum analyses further show that compute-regular and data-parallel workloads are more amenable to LLM optimization, whereas irregular access patterns, dependency-heavy kernels, short-running parallel regions, and CUDA integration constraints remain challenging. 
These results suggest that progress in LLM-based HPC optimization should be measured not only by peak speedups on successful cases, but also by the probability of obtaining correct, robust, and practically useful optimizations across realistic workloads.

\begin{ack}
This work is supported by the National Science Foundation under Award Nos.~2411134 and DOE Genesis Mission Award DE-FOA-0003612.
We acknowledge Anthropic and OpenAI for providing LLM APIs credits used for our experiments. 
We also thank the anonymous reviewers for their valuable feedback.
The authors declare no competing interests.
\end{ack}



{\small
\bibliographystyle{unsrt}
\bibliography{reference}
}


\newpage
\appendix

\let\HPCBenchOrigSection\section
\renewcommand{\section}{\FloatBarrier\HPCBenchOrigSection}

\crefalias{section}{appendix}
\crefalias{subsection}{appendix}  
\crefalias{subsubsection}{appendix}
\Crefname{appendix}{Appendix}{Appendices}
\crefname{appendix}{Appendix}{Appendices}

\section{Full Benchmark Lists}
\label{app_sec:benchmark_list}

This appendix lists all benchmarks used in \HPCBench:
$92$ workloads across $16$ motifs for EX1 and EX2, and $22$ CUDA benchmarks for EX3.

\subsection{Benchmark List for EX1 and EX2}
\label{app_subsec:benchmark_list_cpu}

\Cref{tab:benchmark-list} lists the $92$ CPU/OpenMP benchmarks used by EX1 and EX2, including each benchmark's source suite, motif, structural complexity stratum, and a brief description.

\setlength{\LTpre}{0pt}
\setlength{\LTpost}{0pt}

{\footnotesize
\begin{longtable}{l l c c p{5cm}}
\caption{Benchmark list used in HPC-Bench for EX1 and EX2}
\label{tab:benchmark-list}\\
\toprule
\textbf{Benchmark} & \textbf{Source} & \textbf{Structural Complexity} & \text{Motif} & \textbf{Brief Introduction} \\
\midrule
\endfirsthead

\toprule
Benchmark & Source & Structural Complexity & Motif & Brief Introduction \\
\midrule
\endhead

\midrule
\multicolumn{5}{r}{\emph{Continued on next page}} \\
\endfoot

\bottomrule
\endlastfoot
2mm                  & polybench       & d1                  & DLA            & Two matrix multiplications chain                                          \\
3mm                  & polybench       & d1                  & DLA            & Three matrix multiplications chain                                        \\
adi                  & polybench       & d1                  & SC             & Alternating direction implicit                                            \\
aes                  & machsuite       & d1                  & CE             & Advanced Encryption Standard                                              \\
atax                 & polybench       & d1                  & DLA            & Matrix transpose and vector multiplication                                \\
atmux                & codee           & d1                  & UB             & ATM mux simulation                                                        \\
b+tree               & rodinia         & d3                  & GA             & B+ tree operations                                                        \\
backprop             & rodinia         & d1                  & CML            & Neural network backpropagation                                            \\
basicmath-mibench    & mibench         & d1                  & UB             & Basic math operations                                                     \\
bfs                  & rodinia         & d3                  & GA             & Breadth-first search                                                      \\
bfs-bulk-mach        & machsuite       & d1                  & GA             & BFS with bulk processing                                                  \\
bfs-queue-mach       & machsuite       & d2                  & GA             & BFS with queue                                                            \\
bicg                 & polybench       & d1                  & DLA            & Biconjugate gradients                                                     \\
bilateral            & rodinia         & d1                  & IVP            & Bilateral filter                                                          \\
bitcount-mibench     & mibench         & d1                  & UB             & Bit counting operations                                                   \\
canneal-parsec       & parsec          & d1                  & CML            & Simulated annealing optimization                                          \\
cfd                  & rodinia         & d2                  & CFD            & Computational fluid dynamics solver                                       \\
cholesky             & polybench       & d1                  & DLA            & Cholesky decomposition                                                    \\
correlation          & polybench       & d1                  & STC            & Correlation coefficient                                                   \\
coulomb              & codee           & d1                  & NBM            & Coulombic potential calculation                                           \\
covariance           & polybench       & d1                  & STC            & Covariance matrix                                                         \\
cutcp                & parboil         & d4                  & NBM            & Cutoff Coulombic potential                                                \\
deriche              & polybench       & d1                  & IVP            & Deriche edge detector                                                     \\
dijkstra-mibench     & mibench         & d1                  & GA             & Dijkstra algorithm                                                        \\
doitgen              & polybench       & d1                  & STC            & Multi-resolution analysis                                                 \\
durbin               & polybench       & d1                  & STC            & Durbin's algorithm                                                        \\
fdtd-2d              & polybench       & d1                  & SC             & 2D finite-difference time-domain                                          \\
fft-strided-mach     & machsuite       & d1                  & SM             & FFT with strided access                                                   \\
fft-transpose-mach   & machsuite       & d2                  & SM             & FFT with transpose                                                        \\
floyd-warshall       & polybench       & d1                  & DP             & All-pairs shortest paths                                                  \\
gemm                 & polybench       & d1                  & DLA            & General matrix-matrix multiplication                                      \\
gemm-blocked-mach    & machsuite       & d1                  & DLA            & Blocked matrix multiplication                                             \\
gemm-ncubed-mach     & machsuite       & d1                  & DLA            & Naive O($n^3$) matrix multiplication                                         \\
gemver               & polybench       & d1                  & DLA            & Matrix-vector operations                                                  \\
gesummv              & polybench       & d1                  & DLA            & Scalar-matrix-vector multiplication                                       \\
gramschmidt          & polybench       & d1                  & DLA            & Gram-Schmidt orthogonalization                                            \\
haccmk               & codee           & d1                  & NBM            & HACC short-range force   kernel \\
heartwall            & rodinia         & d4                  & MSC            & Heart wall motion tracking                                                \\
heat-3d              & polybench       & d1                  & SC             & 3D heat equation solver                                                   \\
histo                & parboil         & d2                  & UB             & Image histogram computation     \\
hotspot              & rodinia         & d3                  & SC             & 2D thermal simulation                                                     \\
hotspot3D            & rodinia         & d2                  & SC             & 3D thermal simulation                                                     \\
jacobi-1d            & polybench       & d1                  & SC             & 1D Jacobi iteration                                                       \\
jacobi-2d            & polybench       & d1                  & SC             & 2D Jacobi iteration                                                       \\
kmeans               & rodinia         & d2                  & CML            & K-means clustering                                                        \\
kmp-mach             & machsuite       & d2                  & SS             & Knuth-Morris-Pratt string matching                                        \\
lavaMD               & rodinia         & d1                  & NBM            & Molecular dynamics simulation                                             \\
lbm                  & parboil         & d4                  & CFD            & Lattice Boltzmann method                                                  \\
leukocyte            & rodinia         & d1                  & IVP            & Leukocyte tracking                                                        \\
lu                   & polybench       & d1                  & DLA            & LU decomposition                                                          \\
lud                  & rodinia         & d1                  & DLA            & LU decomposition (variant)                                                \\
ludcmp               & polybench       & d1                  & DLA            & LU decomposition complete                                                 \\
matmul               & codee           & d1                  & DLA            & Basic matrix multiplication                                               \\
md-grid-mach         & machsuite       & d2                  & NBM            & Molecular dynamics with grid                                              \\
md-knn-mach          & machsuite       & d1                  & NBM            & Molecular dynamics with k-NN                                              \\
mri-gridding         & parboil         & d3                  & MSC            & MRI gridding reconstruction                                               \\
mri-q                & parboil         & d1                  & MSC            & MRI Q-space reconstruction                                                \\
mvt                  & polybench       & d1                  & DLA            & Matrix-vector transpose operations                                        \\
myocyte              & rodinia         & d1                  & MSC            & Cardiac myocyte simulation                                                \\
nussinov             & polybench       & d2                  & DP             & RNA secondary structure prediction                                        \\
nw                   & rodinia         & d3                  & DP             & Needleman-Wunsch sequence alignment                                       \\
nw-mach              & machsuite       & d2                  & DP             & Needleman-Wunsch variant                                                  \\
particlefilter       & rodinia         & d2                  & MCM            & Particle filter tracking                                                  \\
pathfinder           & rodinia         & d2                  & DP             & Dynamic programming path finding                                          \\
patricia-mibench     & mibench         & d2                  & GA             & Patricia trie operations                                                  \\
pi                   & codee           & d2                  & MCM            & Monte Carlo $\pi$ estimation                                                  \\
qsort-mibench        & mibench         & d3                  & SS             & Quick sort                                                                \\
sad                  & parboil         & d3                  & IVP            & Sum of absolute differences                                               \\
seidel-2d            & polybench       & d1                  & SC             & 2D Gauss-Seidel iteration                                                 \\
sgemm                & parboil         & d1                  & DLA            & Single-precision matrix multiplication                                    \\
sort-merge-mach      & machsuite       & d1                  & SS             & Merge sort                                                                \\
sort-radix-mach      & machsuite       & d1                  & SS             & Radix sort                                                                \\
spmv                 & parboil         & d1                  & SLA            & Sparse matrix-vector multiplication                                       \\
spmv-crs-mach        & machsuite       & d1                  & SLA            & spmv with CRS format                                                      \\
spmv-ellpack-mach    & machsuite       & d1                  & SLA            & spmv with ELLPACK format                                                  \\
srad\_v1             & rodinia         & d2                  & SC             & Speckle-reducing anisotropic diffusion v1                                 \\
srad\_v2             & rodinia         & d3                  & SC             & Speckle-reducing anisotropic diffusion v2                                 \\
stencil              & parboil         & d1                  & SC             & General stencil operation                                                 \\
stencil-2d-mach      & machsuite       & d1                  & SC             & 2D stencil pattern                                                        \\
stencil-3d-mach      & machsuite       & d1                  & SC             & 3D stencil pattern                                                        \\
streamcluster        & rodinia         & d3                  & CML            & Stream clustering algorithm                                               \\
stringsearch-mibench & mibench         & d1                  & SS             & String search algorithm                                                   \\
susan-c-mibench      & mibench         & d4                  & IVP            & SUSAN corner detection                                                    \\
susan-e-mibench      & mibench         & d4                  & IVP            & SUSAN edge detection                                                      \\
susan-s-mibench      & mibench         & d2                  & IVP            & SUSAN smoothing                                                           \\
symm                 & polybench       & d1                  & DLA            & Symmetric matrix multiplication                                           \\
syr2k                & polybench       & d1                  & DLA            & Symmetric rank-2k update                                                  \\
syrk                 & polybench       & d1                  & DLA            & Symmetric rank-k update                                                   \\
tpacf                & parboil         & d2                  & MCM            & Two-point angular correlation                                             \\
trisolv              & polybench       & d1                  & DLA            & Triangular system solver                                                  \\
trmm                 & polybench       & d1                  & DLA            & Triangular matrix-matrix multiplication                                   \\
viterbi-mach         & machsuite       & d1                  & CE             & Viterbi decoding algorithm                                               
\end{longtable}
}

\subsection{Benchmark List for EX3}
\label{app_subsec:benchmark_list_ex3}

\Cref{tab:benchmark-list-ex3} lists the $22$ Rodinia CUDA benchmarks used by EX3.
This set partially overlaps with the Rodinia benchmarks in the CPU/OpenMP suite, and also includes the five GPU workloads (\texttt{dwt2d}, \texttt{nn}, \texttt{gaussian}, \texttt{huffman}, \texttt{hybridsort}) that are not included in the CPU/OpenMP version.

\setlength{\LTpre}{0pt}
\setlength{\LTpost}{0pt}

{\footnotesize
\begin{longtable}{l l c c p{5cm}}
\caption{Benchmark list used in HPC-Bench for EX3}
\label{tab:benchmark-list-ex3}\\
\toprule
\textbf{Benchmark} & \textbf{Source} & \textbf{Structural Complexity} & \textbf{Motif} & \textbf{Brief Introduction} \\
\midrule
\endfirsthead

\toprule
Benchmark & Source & Structural Complexity & Motif & Brief Introduction \\
\midrule
\endhead

\midrule
\multicolumn{5}{r}{\emph{Continued on next page}} \\
\endfoot

\bottomrule
\endlastfoot

b+tree               & rodinia         & d3                  & GA             & B+ tree operations on GPU                                                 \\
bfs                  & rodinia         & d3                  & GA             & Breadth-first search                                                      \\
bilateral            & rodinia         & d1                  & IVP            & Bilateral filter for image processing                                     \\
cfd                  & rodinia         & d2                  & CFD            & Computational fluid dynamics solver                                       \\
dwt2d                & rodinia         & d2                  & SM             & 2D discrete wavelet transform                                             \\
gaussian             & rodinia         & d1                  & DLA            & Gaussian elimination                                                      \\
heartwall            & rodinia         & d4                  & MSC            & Heart wall motion tracking                                                \\
hotspot              & rodinia         & d3                  & SC             & 2D thermal simulation                                                     \\
hotspot3D            & rodinia         & d2                  & SC             & 3D thermal simulation                                                     \\
huffman              & rodinia         & d2                  & CE             & Huffman encoding/decoding                                                 \\
hybridsort           & rodinia         & d2                  & SS             & Hybrid sorting algorithm                                                  \\
kmeans               & rodinia         & d2                  & CML            & K-means clustering                                                        \\
lavaMD               & rodinia         & d1                  & NBM            & Molecular dynamics simulation                                             \\
leukocyte            & rodinia         & d1                  & IVP            & Leukocyte tracking in video                                               \\
lud                  & rodinia         & d1                  & DLA            & LU decomposition                                                          \\
myocyte              & rodinia         & d1                  & MSC            & Cardiac myocyte simulation                                                \\
nn                   & rodinia         & d1                  & CML            & K-nearest neighbor search                                                 \\
particlefilter       & rodinia         & d2                  & MCM            & Particle filter tracking                                                  \\
pathfinder           & rodinia         & d2                  & DP             & Dynamic programming path finding                                          \\
srad\_v1             & rodinia         & d2                  & SC             & Speckle-reducing anisotropic diffusion v1                                 \\
srad\_v2             & rodinia         & d3                  & SC             & Speckle-reducing anisotropic diffusion v2                                 \\
streamcluster        & rodinia         & d3                  & CML            & Stream clustering algorithm                                               \\

\end{longtable}
}

\subsection{Benchmark Structural Complexity Metadata}
\label{app_subsec:difficulty_metadata}

The structural complexity score in~\Cref{benchmark-suite} is computed from four structural metrics measured on the annotated optimization region: branch count (\texttt{if}, \texttt{case}, ternary operators), lines of code, maximum loop nesting depth, and function count.
Loop nesting depth and function count are extracted using Lizard; LOC and branch count are computed by custom parsers on the annotated region.
Each metric is min-max scaled to $[0,10]$ using the $5$th and $95$th percentiles computed from the $92$ CPU/OpenMP benchmarks before being combined into a weighted composite score.
The labels of CUDA workload's structural complexity use the same scaling statistics, scoring rule, and bucket boundaries as the CPU/OpenMP benchmarks.
The final structural complexity labels are released with the benchmark metadata, as shown in~\Cref{tab:benchmark-list} and~\Cref{tab:benchmark-list-ex3}.

\section{Prompt Templates}
\label{app_sec:prompt_template}

The exact prompts used for EX1, EX2, and EX3 are provided below.
All prompts within a setting share the same template; only the annotated kernel region and the minimal compilation context differ between benchmarks.
Each setting's template specifies the model role, target environment, optimization constraints, and output format using the fixed delimiters $\texttt{<\!<\!<CODE\!>\!>\!>}$ and $\texttt{<\!<\!<END-CODE\!>\!>\!>}$ (see~\Cref{lst:task_format}).

\subsection{EX1: Serial CPU Optimization Prompt}
\label{app_subsec:prompt_ex1}

\label{box:ex1_prompt}
\begin{promptbox}[Prompt Template for EX1]

# System Prompt (Common for EX1 and EX2)
You are a code generation/optimization assistant. Your task is to take a prompt and input code and generate an optimized version of the code.
Your output must strictly follow the output format specified in the task prompt.
The computation environment is a Linux system (Enterprise Linux 8, kernel version 4.18.0-348.7.1) and a dual-socket AMD EPYC 7543 system with 64 physical cores total, SMT disabled.

The available C/C++ compilers are: GCC/G++ v14.2.0 and CLANG/CLANG++ v19.1.3.

# Prompt for EX1 (Serial Optimization)
Provide the optimized C code with serial optimizations for above input code, ensuring that none of the existing functions or header files are removed, and no new functions or print statements are added.
Wrap the generated optimized C code between <<<CODE>>> and <<<END-CODE>>>. Then provide an explanation of why this optimization was made, and wrap the explanation between <<<NOTE>>> and <<<END-NOTE>>>.
Ensure that the optimized code contains only compilable source code, and the explanation contains only plain text or code snippets as needed.
Do not emit any extra commentary before or after these markers.
Example format:
<<<CODE>>>
static void foo() {
  // optimized code body
}
<<<END-CODE>>>
<<<NOTE>>>
Explain the optimization rationale here.
<<<END-NOTE>>>

\end{promptbox}

\subsection{EX2: OpenMP CPU Parallelization Prompt}
\label{app_subsec:prompt_ex2}

The EX2 prompt explicitly forbids \texttt{omp\_set\_num\_threads} and any modification of the timing instrumentation in the source;
thread counts are controlled externally via the \texttt{OMP\_NUM\_THREADS} environment variable.

\label{box:ex2_prompt}
\begin{promptbox}[Prompt Template for EX2]

# System Prompt (Common for EX1 and EX2)
You are a code generation/optimization assistant. Your task is to take a prompt and input code and generate an optimized version of the code.
Your output must strictly follow the output format specified in the task prompt.
The computation environment is a Linux system (Enterprise Linux 8, kernel version 4.18.0-348.7.1) and a dual-socket AMD EPYC 7543 system with 64 physical cores total, SMT disabled.
The available C/C++ compilers are: GCC/G++ v14.2.0 and CLANG/CLANG++ v19.1.3.

# Prompt for EX2 (Serial + Parallel Optimization with OpenMP)
Provide the optimized C code with both serial and parallel optimizations, ensuring that none of the existing functions or header files are removed, and no new functions or print statements are added.
The file already contains an `_OPENMP`-guarded `#include <omp.h>`; keep it intact and feel free to use OpenMP pragmas/API directly, but never insert `omp_set_num_threads` (thread counts come from the runtime via `OMP_NUM_THREADS`).
Do not modify, relocate, or duplicate any time-measurement instrumentation (`clock_gettime`, `omp_get_wtime`, kernel-time accumulators, or related globals); if you restructure control flow, ensure the function still reaches the existing end-of-function timing accumulation (avoid early returns by storing results first).
Avoid adding printf/logging inside parallel regions so multi-threaded output stays deterministic.
Wrap the generated optimized C code between <<<CODE>>> and <<<END-CODE>>>. Then provide an explanation of why this optimization was made, and wrap the explanation between <<<NOTE>>> and <<<END-NOTE>>>.
Ensure that the optimized code contains only compilable source code, and the explanation contains only plain text or code snippets as needed.
Do not emit anything outside those markers.
Example:
<<<CODE>>>
/* optimized code */
<<<END-CODE>>>
<<<NOTE>>>
Explain the optimization rationale here.
<<<END-NOTE>>>

\end{promptbox}

\subsection{EX3: CUDA GPU Optimization Prompt}
\label{app_subsec:prompt_ex3}

The EX3 prompt specifies the CUDA / \texttt{nvcc} environment and constrains the LLM to preserve the host-device memory layout, the external interfaces, and the timing/instrumentation hooks of the original benchmark.

\label{box:ex3_prompt}
\begin{promptbox}[Prompt Template for EX3]

# System Prompt for EX3 (CUDA Main Computational Kernel Optimization)
You are a CUDA kernel optimization expert.
Your output must strictly follow the output format specified in the task prompt.
Work on Linux (RHEL 8, kernel 4.18.0-348.7.1) with an AMD EPYC 7543 host CPU.
Host code compiles with GCC/G++ 14.2 or Clang/Clang++ 19.1.
All GPU code must compile with NVIDIA nvcc 12.4 (CUDA Toolkit 12.4, driver 550.99), targeting NVIDIA A100-class GPUs.

Your task is to optimize the annotated CUDA computational kernel region.
The <<<CODE>>> block must contain only compilable CUDA C/C++ source code.
Do not emit extra commentary outside the requested markers.

# Prompt for EX3 (CUDA Main Computational Kernel Optimization)

Target file: <benchmark-specific CUDA source file>.
Optimize the annotated CUDA kernel region and any benchmark-provided wrapper code included in the prompt.
Do not change external interfaces, host--device memory layouts, kernel launch semantics, input/output parameter lists, or timing/instrumentation hooks outside the annotated region.

Goals:
1. Improve memory efficiency: maximize coalesced global-memory loads/stores, reduce redundant memory traffic, use shared memory when it is beneficial and safe, and add __restrict__ qualifiers only when legally valid.
2. Improve parallel execution efficiency: reduce unnecessary thread divergence, improve occupancy when possible, avoid excessive register/shared-memory usage, and use warp-level primitives only when they preserve correctness.
3. Preserve synchronization and memory semantics: keep required __syncthreads() barriers, do not remove boundary checks unless they are provably redundant, and do not introduce races or out-of-bounds accesses.
4. Preserve correctness and integration: do not change host-side allocation, data transfer, logging, timing code, or benchmark I/O behavior. Do not insert cudaDeviceSynchronize unless it is required for correctness and does not interfere with existing timing instrumentation.
5. Preserve output format: comments are allowed inside the code block, but the generated code must be directly compilable after being inserted back into the original benchmark source.

Wrap the optimized CUDA code between <<<CODE>>> and <<<END-CODE>>>.
If you provide rationale, wrap it between <<<NOTE>>> and <<<END-NOTE>>> with plain text only.
The evaluator will extract and compile only the <<<CODE>>> block.

Example:
<<<CODE>>>
__global__ void optimized_kernel(...) {
    // optimized CUDA kernel body
}
<<<END-CODE>>>
<<<NOTE>>>
Briefly explain the main optimization idea.
<<<END-NOTE>>>
\end{promptbox}

\section{Reproducibility Details}
\label{app_sec:reproducibility}

Compilation flags, input construction, and timing protocols for each setting are detailed below.

\subsection{Compilation and Link Flags}
\label{app_subsec:compilation_flags}

\autoref{tab:ex1_flags} reports compilation and link flags for EX1 (Serial CPU).
EX2 (\autoref{tab:ex2_flags}) uses the same flags with \texttt{-fopenmp} added to compilation and linking; thread counts are set through \texttt{OMP\_NUM\_THREADS}.
EX3 (\autoref{tab:ex3_flags}) lists the CUDA additions for each benchmark on top of a common \texttt{nvcc} 12.4 base configuration with \texttt{--generate-line-info}, host \texttt{-O3}, and link flags \texttt{-lcudart -lnvToolsExt}.

\newcommand{\cc}[1]{\texttt{#1}}

{
\rowcolors{2}{gray!8}{white}


\begin{longtable}{p{3.8cm} p{1.6cm} p{0.8cm} p{5.2cm} p{3cm}}
\caption{EX1 (Serial CPU) compilation flags. Compiler: GCC/G++ 14.2.0, Enterprise Linux 8. \cc{haccmk} uses \cc{-O2} for driver, \cc{-O3} for kernel.}\label{tab:ex1_flags}\\
\toprule
\textbf{Benchmark} & \textbf{Suite} & \textbf{CC} & \textbf{Compilation Flags} & \textbf{Link Flags} \\
\midrule
\endfirsthead
\toprule
\textbf{Benchmark} & \textbf{Suite} & \textbf{CC} & \textbf{Compilation Flags} & \textbf{Link Flags} \\
\midrule
\endhead
\midrule
\multicolumn{5}{r}{\footnotesize\itshape Continued on next page\ldots}\\
\endfoot
\bottomrule
\endlastfoot

\cc{2mm}              & PolyBench & gcc & \cc{-O3}  & \cc{-lm} \\
\cc{3mm}              & PolyBench & gcc & \cc{-O3}  & \cc{-lm} \\
\cc{adi}              & PolyBench & gcc & \cc{-O3}  & \cc{-lm} \\
\cc{atax}             & PolyBench & gcc & \cc{-O3}  & \cc{-lm} \\
\cc{bicg}             & PolyBench & gcc & \cc{-O3}  & \cc{-lm} \\
\cc{cholesky}         & PolyBench & gcc & \cc{-O3}  & \cc{-lm} \\
\cc{correlation}      & PolyBench & gcc & \cc{-O3}  & \cc{-lm} \\
\cc{covariance}       & PolyBench & gcc & \cc{-O3}  & \cc{-lm} \\
\cc{deriche}          & PolyBench & gcc & \cc{-O3}  & \cc{-lm} \\
\cc{doitgen}          & PolyBench & gcc & \cc{-O3}  & \cc{-lm} \\
\cc{durbin}           & PolyBench & gcc & \cc{-O3}  & \cc{-lm} \\
\cc{fdtd-2d}          & PolyBench & gcc & \cc{-O3}  & \cc{-lm} \\
\cc{floyd-warshall}   & PolyBench & gcc & \cc{-O3}  & \cc{-lm} \\
\cc{gemm}             & PolyBench & gcc & \cc{-O3}  & \cc{-lm} \\
\cc{gemver}           & PolyBench & gcc & \cc{-O3}  & \cc{-lm} \\
\cc{gesummv}          & PolyBench & gcc & \cc{-O3}  & \cc{-lm} \\
\cc{gramschmidt}      & PolyBench & gcc & \cc{-O3}  & \cc{-lm} \\
\cc{heat-3d}          & PolyBench & gcc & \cc{-O3}  & \cc{-lm} \\
\cc{jacobi-1d}        & PolyBench & gcc & \cc{-O3}  & \cc{-lm} \\
\cc{jacobi-2d}        & PolyBench & gcc & \cc{-O3}  & \cc{-lm} \\
\cc{lu}               & PolyBench & gcc & \cc{-O3}  & \cc{-lm} \\
\cc{ludcmp}           & PolyBench & gcc & \cc{-O3}  & \cc{-lm} \\
\cc{mvt}              & PolyBench & gcc & \cc{-O3}  & \cc{-lm} \\
\cc{nussinov}         & PolyBench & gcc & \cc{-O3}  & \cc{-lm} \\
\cc{seidel-2d}        & PolyBench & gcc & \cc{-O3}  & \cc{-lm} \\
\cc{symm}             & PolyBench & gcc & \cc{-O3}  & \cc{-lm} \\
\cc{syr2k}            & PolyBench & gcc & \cc{-O3}  & \cc{-lm} \\
\cc{syrk}             & PolyBench & gcc & \cc{-O3}  & \cc{-lm} \\
\cc{trisolv}          & PolyBench & gcc & \cc{-O3}  & \cc{-lm} \\
\cc{trmm}             & PolyBench & gcc & \cc{-O3}  & \cc{-lm} \\
\midrule

\cc{backprop}         & Rodinia & gcc & \cc{-O3}              & \cc{-lm} \\
\cc{bfs}              & Rodinia & g++ & \cc{-O3}              & \cc{-lm} \\
\cc{bilateral}        & Rodinia & g++ & \cc{-O3}              & \cc{-lm} \\
\cc{b+tree}           & Rodinia & gcc & \cc{-O3 -w}           & \cc{-lm} \\
\cc{cfd}              & Rodinia & g++ & \cc{-O3}              & \cc{-lm} \\
\cc{heartwall}        & Rodinia & gcc & \cc{-O3}              & \cc{-lm} \\
\cc{hotspot}          & Rodinia & g++ & \cc{-O3 -w}           & \cc{-lm} \\
\cc{hotspot3D}        & Rodinia & gcc & \cc{-O3}              & \cc{-lm} \\
\cc{kmeans}           & Rodinia & gcc & \cc{-O3}              & \cc{-lm} \\
\cc{lavaMD}           & Rodinia & gcc & \cc{-O3}              & \cc{-lm} \\
\cc{leukocyte}        & Rodinia & gcc & \cc{-O3}              & \cc{-lm} \\
\cc{lud}              & Rodinia & gcc & \cc{-O3}              & \cc{-lm} \\
\cc{myocyte}          & Rodinia & gcc & \cc{-O3}              & \cc{-lm} \\
\cc{nw}               & Rodinia & g++ & \cc{-O3}              & \cc{-lm} \\
\cc{particlefilter}   & Rodinia & gcc & \cc{-O3}              & \cc{-lm} \\
\cc{pathfinder}       & Rodinia & g++ & \cc{-O3}              & \cc{-lm} \\
\cc{srad\_v1}         & Rodinia & gcc & \cc{-O3}              & \cc{-lm} \\
\cc{srad\_v2}         & Rodinia & g++ & \cc{-O3}              & \cc{-lm} \\
\cc{streamcluster}    & Rodinia & g++ & \cc{-O3}              & \cc{-lpthread -lm} \\
\midrule

\cc{cutcp}            & Parboil & gcc & \cc{-O3}                                 & \cc{-lm -lpthread} \\
\cc{histo}            & Parboil & gcc & \cc{-O3}                                 & \cc{-lm -lpthread} \\
\cc{lbm}              & Parboil & gcc & \cc{-O3}                                 & \cc{-lm -lpthread} \\
\cc{mri-gridding}     & Parboil & gcc & \cc{-O3 -w}                              & \cc{-lm -lpthread} \\
\cc{mri-q}            & Parboil & g++ & \cc{-O3 -ffast-math -funroll-loops -w}   & \cc{-lm} \\
\cc{sad}              & Parboil & gcc & \cc{-O3}                                 & \cc{-lm -lpthread} \\
\cc{sgemm}            & Parboil & g++ & \cc{-O3 -w}                              & \cc{-lm -lstdc++} \\
\cc{spmv}             & Parboil & gcc & \cc{-O3}                                 & \cc{-lm -lpthread} \\
\cc{stencil}          & Parboil & gcc & \cc{-O3}                                 & \cc{-lm -lpthread} \\
\cc{tpacf}            & Parboil & gcc & \cc{-O3}                                 & \cc{-lm -lpthread} \\
\midrule

\cc{aes}              & MachSuite & gcc & \cc{-O3 -Wall -Wno-unused-label} & --- \\
\cc{bfs-bulk-mach}    & MachSuite & gcc & \cc{-O3 -Wall -Wno-unused-label} & --- \\
\cc{bfs-queue-mach}   & MachSuite & gcc & \cc{-O3 -Wall -Wno-unused-label} & --- \\
\cc{fft-strided-mach} & MachSuite & gcc & \cc{-O3 -Wall -Wno-unused-label} & --- \\
\cc{fft-transpose-mach} & MachSuite & gcc & \cc{-O3 -Wall -Wno-unused-label} & --- \\
\cc{gemm-blocked-mach}  & MachSuite & gcc & \cc{-O3 -Wall -Wno-unused-label} & --- \\
\cc{gemm-ncubed-mach}   & MachSuite & gcc & \cc{-O3 -Wall -Wno-unused-label} & --- \\
\cc{kmp-mach}          & MachSuite & gcc & \cc{-O3 -Wall -Wno-unused-label} & --- \\
\cc{md-grid-mach}      & MachSuite & gcc & \cc{-O3 -Wall -Wno-unused-label} & --- \\
\cc{md-knn-mach}       & MachSuite & gcc & \cc{-O3 -Wall -Wno-unused-label} & --- \\
\cc{nw-mach}           & MachSuite & gcc & \cc{-O3 -Wall -Wno-unused-label} & --- \\
\cc{sort-merge-mach}   & MachSuite & gcc & \cc{-O3 -Wall -Wno-unused-label} & --- \\
\cc{sort-radix-mach}   & MachSuite & gcc & \cc{-O3 -Wall -Wno-unused-label} & --- \\
\cc{spmv-crs-mach}     & MachSuite & gcc & \cc{-O3 -Wall -Wno-unused-label} & --- \\
\cc{spmv-ellpack-mach}  & MachSuite & gcc & \cc{-O3 -Wall -Wno-unused-label} & --- \\
\cc{stencil-2d-mach}   & MachSuite & gcc & \cc{-O3 -Wall -Wno-unused-label} & --- \\
\cc{stencil-3d-mach}   & MachSuite & gcc & \cc{-O3 -Wall -Wno-unused-label} & --- \\
\cc{viterbi-mach}      & MachSuite & gcc & \cc{-O3 -Wall -Wno-unused-label} & --- \\
\midrule

\cc{basicmath-mibench}     & MiBench & gcc & \cc{-O3 -w}                & \cc{-lm} \\
\cc{bitcount-mibench}      & MiBench & gcc & \cc{-O3}                   & --- \\
\cc{dijkstra-mibench}      & MiBench & gcc & \cc{-O3 -w -std=gnu89}    & --- \\
\cc{patricia-mibench}      & MiBench & gcc & \cc{-O3 -w -std=gnu89}    & --- \\
\cc{qsort-mibench}         & MiBench & gcc & \cc{-O3}                   & \cc{-lm} \\
\cc{stringsearch-mibench}  & MiBench & gcc & \cc{-O3 -w}               & --- \\
\cc{susan-c-mibench}       & MiBench & gcc & \cc{-O3 -w}               & \cc{-lm} \\
\cc{susan-e-mibench}       & MiBench & gcc & \cc{-O3 -w}               & \cc{-lm} \\
\cc{susan-s-mibench}       & MiBench & gcc & \cc{-O3 -w}               & \cc{-lm} \\
\midrule

\cc{canneal-parsec}         & PARSEC & g++ & \cc{-O3 -std=c++11}       & \cc{-lm} \\
\midrule

\cc{atmux}     & CODEE & gcc & \cc{-std=c99 -O3}                       & \cc{-lm -lrt} \\
\cc{coulomb}   & CODEE & gcc & \cc{-O3}                                 & \cc{-lm} \\
\cc{haccmk}    & CODEE & gcc & \cc{-O2} (main) / \cc{-O3} (kernel) & \cc{-lm} \\
\cc{matmul}    & CODEE & gcc & \cc{-O3}                                 & --- \\
\cc{pi}        & CODEE & gcc & \cc{-O3}                                 & \cc{-lm} \\

\end{longtable}


\begin{table}[ht]
\centering
\caption{EX2 (OpenMP CPU) compilation flags. EX2 uses identical flags to EX1 (\autoref{tab:ex1_flags}) with \cc{-fopenmp} added to both compilation and link flags for all 92 benchmarks. Thread counts are controlled at runtime via \cc{OMP\_NUM\_THREADS}.}\label{tab:ex2_flags}
\end{table}


\begin{longtable}{p{3.2cm} p{4.5cm} p{4.5cm}}
\caption{EX3 (CUDA GPU) compilation flags. Common base: \cc{nvcc} 12.4, \cc{--generate-line-info}, host \cc{-O3}, link \cc{-lcudart -lnvToolsExt}. Only per-benchmark additions are listed.}\label{tab:ex3_flags}\\
\toprule
\textbf{Benchmark} & \textbf{Additional NVCCFLAGS} & \textbf{Additional Link Flags} \\
\midrule
\endfirsthead
\toprule
\textbf{Benchmark} & \textbf{Additional NVCCFLAGS} & \textbf{Additional Link Flags} \\
\midrule
\endhead
\bottomrule
\endlastfoot

\cc{bfs}             & ---            & --- \\
\cc{bilateral}       & ---            & --- \\
\cc{b+tree}          & ---            & \cc{-lm -lstdc++} \\
\cc{cfd}             & ---            & --- \\
\cc{dwt2d}           & \cc{-DOUTPUT -w} & --- \\
\cc{gaussian}        & ---            & --- \\
\cc{heartwall}       & ---            & --- \\
\cc{hotspot}         & ---            & --- \\
\cc{hotspot3D}       & ---            & --- \\
\cc{huffman}         & \cc{-w}        & --- \\
\cc{hybridsort}      & \cc{-w}        & --- \\
\cc{kmeans}          & \cc{-w}        & --- \\
\cc{lavaMD}          & \cc{-w}        & --- \\
\cc{leukocyte}       & \cc{-w}        & \cc{-lm -lstdc++} \\
\cc{lud}             & \cc{-w}        & --- \\
\cc{myocyte}         & \cc{-w}        & \cc{-lm} \\
\cc{nn}              & ---            & --- \\
\cc{particlefilter}  & \cc{-w}        & \cc{-lm} \\
\cc{pathfinder}      & \cc{-w}        & --- \\
\cc{srad\_v1}        & \cc{-w}        & --- \\
\cc{srad\_v2}        & \cc{-w}        & --- \\
\cc{streamcluster}   & \cc{-w}        & --- \\

\end{longtable}

}  

\subsection{Input Construction and Correctness Checking}
\label{app_subsec:inputs}

We use the original benchmark inputs where available and manually construct the missing sizes among the five standardized scales (\emph{mini}, \emph{small}, \emph{medium}, \emph{large}, \emph{extra-large}).
LLMs are not used for input generation or correctness checking.
Each valid input has a reference produced by the current baseline program.
Each candidate is checked against the reference using benchmark-specific comparisons: 
numerical tolerances for numeric results, exact matching for text, and binary comparison for image or binary outputs.
Only candidates that pass the comparison receive correctness credit; 
skipped checks, failed checks, crashes, and timeouts count as incorrect.

\subsection{Hardware, Timing, and Runtime Measurement}
\label{app_subsec:runtime_protocol}

\paragraph{Hardware.}
EX1 and EX2 run on a dual-socket AMD EPYC 7543 with 64 physical cores total (32 per socket), simultaneous multithreading disabled, under Enterprise Linux 8.
EX3 runs on an NVIDIA A100 GPU using CUDA Toolkit 12.4 / \texttt{nvcc} 12.4.
Host code in EX3 uses the same GCC toolchain as EX1/EX2.

\paragraph{Timing Protocol.}
For each benchmark and configuration, we perform warmup iterations followed by repeated timed trials and report the mean runtime.
Warmup and trial counts are selected adaptively from a preliminary run to accommodate different runtime scales.

\paragraph{Reported Quantities.}
We report three quantities:
\begin{enumerate}
    \item \textbf{Kernel Runtime} is the runtime of the annotated optimization region and is the primary signal for all metrics in the main paper;
    \item \textbf{Total Program Runtime} is the full benchmark execution time (input setup + allocation + I/O + kernel + cleanup), used for comparison in~\Cref{app_subsec:total_runtime_metrics};
    \item \textbf{Evaluation Cost} is the API cost for generation, reported in~\Cref{app_sec:budget}.
\end{enumerate}

\paragraph{EX2 Thread Control.}
EX2 thread counts are set at runtime through \texttt{OMP\_NUM\_THREADS}, with source-level overrides prohibited.
The main paper reports EX2 results at $n=16$ threads; the full sweep over $n\in\{1,2,8,16,32\}$ is in~\Cref{app_subsec:thread_sensitivity}.

\paragraph{Compute Resource Usage.}
Across all evaluated models and input sizes, the reported experiments required approximately \textbf{$300$ CPU node-hours} for EX1 (Serial), \textbf{$1200$ CPU node-hours} for EX2 (OpenMP), and \textbf{$100$ GPU-hours} for EX3 (CUDA).
These estimates cover compilation, correctness verification, and performance measurement.

\section{Full Metric Definitions}
\label{app_sec:metric_full}

This section provides metric notation, auxiliary statistics, and the aggregation protocol for $\mathrm{Fast}@k$ and $\mathrm{Speedup}@k$.

\subsection{Notation}
\label{app_subsec:metric_notation}

\autoref{tab:notation} lists the symbols used in the metric definitions.

\begin{table}[H]
\centering
\small
\setlength{\tabcolsep}{4pt}
\caption{Notation used in the metric definitions. All metrics are computed per benchmark and then macro-averaged across the relevant stratum.}
\label{tab:notation}
\begin{tabular}{@{}p{0.20\linewidth}p{0.74\linewidth}@{}}
\toprule
\textbf{Symbol} & \textbf{Description} \\
\midrule
$b\in\mathcal{B}$ & Benchmark index \\
$\ell\in\mathcal{L}$ & Model index (Claude Sonnet 4.5, GPT-5.1, Qwen3-Coder) \\
$d(b)\in\{d_1,d_2,d_3,d_4\}$ & Structural complexity stratum of benchmark $b$ \\
$m(b)\in\mathcal{M}$ & Computational motif of benchmark $b$ \\
$j\in\{1,\dots,N\}$ & Candidate index (we use $N=10$) \\
$N$ & Number of generated candidates per (benchmark, model) \\
$k$ & Sampling budget; $k\le N$ (we use $k=3$ unless noted) \\
$n$ & Thread count (relevant only for EX2; $n\in\{1,2,8,16,32\}$) \\
$t\in\{\mathrm{kernel},\mathrm{total}\}$ & Runtime measurement type \\
$\hat{T}^{(t)}_{b,n}$ & Baseline runtime of benchmark $b$ under runtime type $t$ and threads $n$ \\
$T^{(t)}_{b,j,n}$ & Runtime of candidate $j$ under runtime type $t$ and threads $n$ \\
$\mathrm{correct}_{b,j}\in\{0,1\}$ & Correctness indicator (matches reference output: $1$, otherwise $0$) \\
$S^{(t)}_{b,j,n}$ & Correctness-gated speedup (defined below) \\
$p_{\mathrm{thr}}$ & Speedup threshold, $p_{\mathrm{thr}}>0$; $p_{\mathrm{thr}}=1.0$ requires no slowdown, while $p_{\mathrm{thr}}>1$ imposes a minimum speedup \\
\bottomrule
\end{tabular}
\end{table}

\subsection{Correctness-Gated Speedup}
\label{app_subsec:correctness_gated_speedup}

Candidates that fail compilation, execution, or correctness checking receive zero speedup:
\begin{equation}
S^{(t)}_{b,j,n}
\;=\;
\Big(\hat{T}^{(t)}_{b,n}\,/\,T^{(t)}_{b,j,n}\Big)
\,\cdot\,
\mathbf{1}\!\big\{\mathrm{correct}_{b,j}=1\big\}.
\end{equation}
Failed candidates remain in the sample pool of size $N$ and contribute zero to all sampling-aware metrics.
This convention makes $\mathrm{Speedup}@k$ monotonically non-increasing in $p_{\mathrm{thr}}$ (\Cref{subsec:evaluation-metrics}).

\subsection{Auxiliary Metrics}
\label{app_subsec:auxiliary_metrics}

We report three auxiliary statistics.

\paragraph{Pass Rate.}
The fraction of generated candidates that are functionally correct is
\begin{equation}
\mathrm{pass}(b)\;=\;\frac{1}{N}\sum_{j=1}^{N}\mathrm{correct}_{b,j}.
\end{equation}

\paragraph{Best-of-$N$ Speedup (Correct-only).}
The maximum correctness-gated speedup over all $N$ candidates is
\begin{equation}
\mathrm{best}^{(t)}_{n}(b)\;=\;\max_{j\in\{1,\dots,N\}}\,S^{(t)}_{b,j,n}.
\end{equation}

\paragraph{Mean Runtime Over Correct Candidates.}
\begin{equation}
\mathrm{MeanT}^{(t)}_{n}(b)
\;=\;
\frac{\sum_{j=1}^{N}T^{(t)}_{b,j,n}\cdot\mathrm{correct}_{b,j}}{\sum_{j=1}^{N}\mathrm{correct}_{b,j}},
\end{equation}
which is undefined if no correct candidate is generated.

Aggregated values for these auxiliary statistics at the EX level are reported in~\Cref{tab:aux_ex1}, \Cref{tab:aux_ex2}, and \Cref{tab:aux_ex3}.

\begin{table}[H]
\centering
\captionsetup{justification=centering}
\caption{Auxiliary metrics for EX1 (Serial CPU optimization), averaged over all benchmarks at the medium dataset size.
Pass rate, best-of-$N$ speedup, and MeanT are first computed per benchmark and then macro-averaged.
MeanT reports the average runtime over correct candidates only.}
\label{tab:aux_ex1}
\small
\begin{tabular}{lccc}
\toprule
Model & Pass rate & Best-of-$N$ speedup & MeanT (s) \\
\midrule
GPT-5.1 & 0.76 & 2.16 & 1.88 \\
Claude  & 0.80 & 2.13 & 2.10 \\
Qwen    & 0.74 & 1.44 & 2.00 \\
\bottomrule
\end{tabular}
\end{table}

\begin{table}[H]
\centering
\captionsetup{justification=centering}
\caption{Auxiliary metrics for EX2 (OpenMP CPU parallelization), averaged over all benchmarks at the medium dataset size.
Pass rate, best-of-$N$ speedup, and MeanT are first computed per benchmark and then macro-averaged.
MeanT reports the average runtime over correct candidates only.}
\label{tab:aux_ex2}
\small
\begin{tabular}{lccc}
\toprule
Model & Pass rate & Best-of-$N$ speedup & MeanT (s) \\
\midrule
GPT-5.1 & 0.59 & 3.65 & 1.18 \\
Claude  & 0.65 & 3.49 & 1.54 \\
Qwen    & 0.59 & 2.68 & 1.48 \\
\bottomrule
\end{tabular}
\end{table}

\begin{table}[H]
\centering
\captionsetup{justification=centering}
\caption{Auxiliary metrics for EX3 (CUDA GPU optimization), averaged over all benchmarks at the medium dataset size.
Pass rate, best-of-$N$ speedup, and MeanT are first computed per benchmark and then macro-averaged.
MeanT reports the average runtime over correct candidates only.}
\label{tab:aux_ex3}
\small
\begin{tabular}{lccc}
\toprule
Model & Pass rate & Best-of-$N$ speedup & MeanT (s) \\
\midrule
GPT-5.1 & 0.36 & 3.53 & 0.12 \\
Claude  & 0.42 & 2.76 & 0.30 \\
Qwen    & 0.38 & 2.49 & 0.44 \\
\bottomrule
\end{tabular}
\end{table}

\subsection{Aggregation and Stratified Reporting}
\label{app_subsec:aggregation}

Unless otherwise stated, aggregate values are macro-averaged across benchmarks within a stratum.
Given a grouping criterion $g(b)\in\{m(b),d(b),\text{dataset size},\dots\}$, define $\mathcal{B}_{g=c}=\{b\in\mathcal{B}\,:\,g(b)=c\}$.
For a metric $\mu(b)\in\{\mathrm{pass}, \mathrm{Fast}@k, \mathrm{Speedup}@k, \mathrm{best}^{(t)}_n, \mathrm{MeanT}^{(t)}_n\}$ stratified by $g$,
\begin{equation}
\mu\big|_{g=c}\;=\;\frac{1}{|\mathcal{B}_{g=c}|}\sum_{b\in\mathcal{B}_{g=c}}\mu(b).
\end{equation}
For figures aggregated over input sizes, we compute the metric at each of the five scales (mini, small, medium, large, extra-large) and then average across scales.
For sensitivity analyses with respect to $k$, results are reported on the medium dataset unless otherwise noted.

\section{Motif-Level Characterization}
\label{app_sec:motif_characterization}

\autoref{tab:motif_characterization} summarizes the $16$ motifs in \HPCBench by their dominant performance bottleneck, representative access and dependency patterns, parallelization structure, common optimization opportunities, and representative kernels.

\begin{table}[H]
\centering
\captionsetup{labelsep=period, justification=centering}
\caption{Motif-level HPC characterization. For each computational motif we summarize the dominant bottleneck, representative access/dependency pattern, parallelization structure, typical optimization opportunities, and representative kernels from the benchmark suite.}
\label{tab:motif_characterization}
\resizebox{\textwidth}{!}{%
\begin{tabular}{lrcllll}
\toprule
\textbf{Motif} & \textbf{\#B} & \textbf{Dominant Bottleneck} & \textbf{Representative Access / Dependency Pattern} & \textbf{Parallelization Structure} & \textbf{Representative Optimization Opportunities} & \textbf{Representative HPC-Bench Kernels} \\
\midrule
Cryptography/Encoding (CE)           &  2 & Compute-bound            & Sequential, bit-level                  & Sequential (block/state dep.)        & Lookup tables, SIMD                              & aes, viterbi-mach \\
Comp.\ Fluid Dynamics (CFD)          &  2 & Bandwidth/Compute-bound  & Structured grid                        & Data-parallel                        & Cache blocking, SIMD                             & cfd, lbm \\
Clustering/ML (CML)                  &  4 & Compute/Memory-bound     & Iterative convergence, random access   & Data-parallel / barrier-synchronized & Convergence optimization, reduction              & kmeans, backprop, streamcluster \\
Dense Linear Algebra (DLA)           & 22 & Compute-bound            & Regular (GEMM, matvec) to triangular (factor.) & Data-parallel to level-sequential    & Loop tiling/reorder, blocked algorithms          & gemm, cholesky, trisolv \\
Dynamic Programming (DP)             &  5 & Compute-bound            & Wavefront, anti-diagonal               & Wavefront                            & Memoization, loop restructuring                  & floyd-warshall, nw, nw-mach \\
Graph Algorithms (GA)                &  6 & Memory-bound             & Irregular, pointer-chasing             & Task-parallel                        & Queue optimization, frontier-based               & bfs, b+tree, dijkstra-mibench \\
Image/Video Processing (IVP)         &  7 & Memory-bound             & 2D spatial / separable recursive       & Data-parallel                        & Window optimization, vectorization               & bilateral, deriche, sad \\
Monte Carlo Methods (MCM)            &  3 & Compute-bound            & Random, independent samples            & Embarrassingly parallel              & RNG optimization, reduction                      & pi, particlefilter, tpacf \\
Medical/Scientific Comp.\ (MSC)      &  4 & Compute/Memory-bound     & ODE time-stepping / grid / scatter-gather & Sequential to data-parallel       & Loop restructuring, vectorization                & myocyte, mri-q, mri-gridding \\
N-body Methods (NBM)                 &  6 & Compute-bound            & Pairwise, $O(N^2)$                     & Embarrassingly parallel              & Cutoff radius, spatial decomposition             & lavaMD, coulomb, cutcp \\
Stencil/Scientific Comp.\ (SC)       & 13 & Bandwidth/Compute-bound  & Regular, neighbor stencil              & Data-parallel (impl.\ solver variants) & Temporal blocking, tiling, vectorization       & jacobi-2d, hotspot, adi \\
Sparse Linear Algebra (SLA)          &  3 & Memory-bound             & Irregular (indirect indexing)          & Row-parallel (index-constrained)   & Format conversion, prefetch, layout opt.         & spmv, spmv-crs-mach, spmv-ellpack-mach \\
Spectral Methods (SM)                &  2 & Memory-bound             & Strided, transpose                     & Butterfly stages                     & Memory layout optimization                       & fft-strided-mach, fft-transpose-mach \\
Sorting/Search (SS)                  &  5 & Memory-bound             & Data-dependent                         & Sequential to task-level             & Branch reduction, restructuring                  & sort-radix-mach, kmp-mach, qsort-mibench \\
Statistical Comp.\ (STC)             &  4 & Reduction-dominated      & Regular, reduction                     & Data-parallel                        & Loop fusion, reduction optimization              & correlation, covariance, doitgen \\
Utility Benchmarks (UB)              &  4 & Compute/Memory-bound     & Scalar / random-scatter / indirect     & Sequential to reduction-based        & Intrinsics, thread-local reduction               & bitcount-mibench, basicmath-mibench, atmux \\
\bottomrule
\end{tabular}%
}
\end{table}

\section{Additional Evaluation Results}
\label{app_sec:additional_results}

This section provides additional analysis for RQ1, RQ2, and RQ3.
It examines input-size effects, speedup distributions, motif and structural complexity differences, sampling budgets, OpenMP thread counts, and total runtime.
Unless otherwise stated, results use kernel runtime and the correctness gating defined in~\Cref{app_sec:metric_full}.

\subsection{Input-Size Sensitivity}
\label{app_subsec:input_size_sensitivity}

\begin{figure*}[t]
  \centering
  \includegraphics[width=0.95\textwidth]{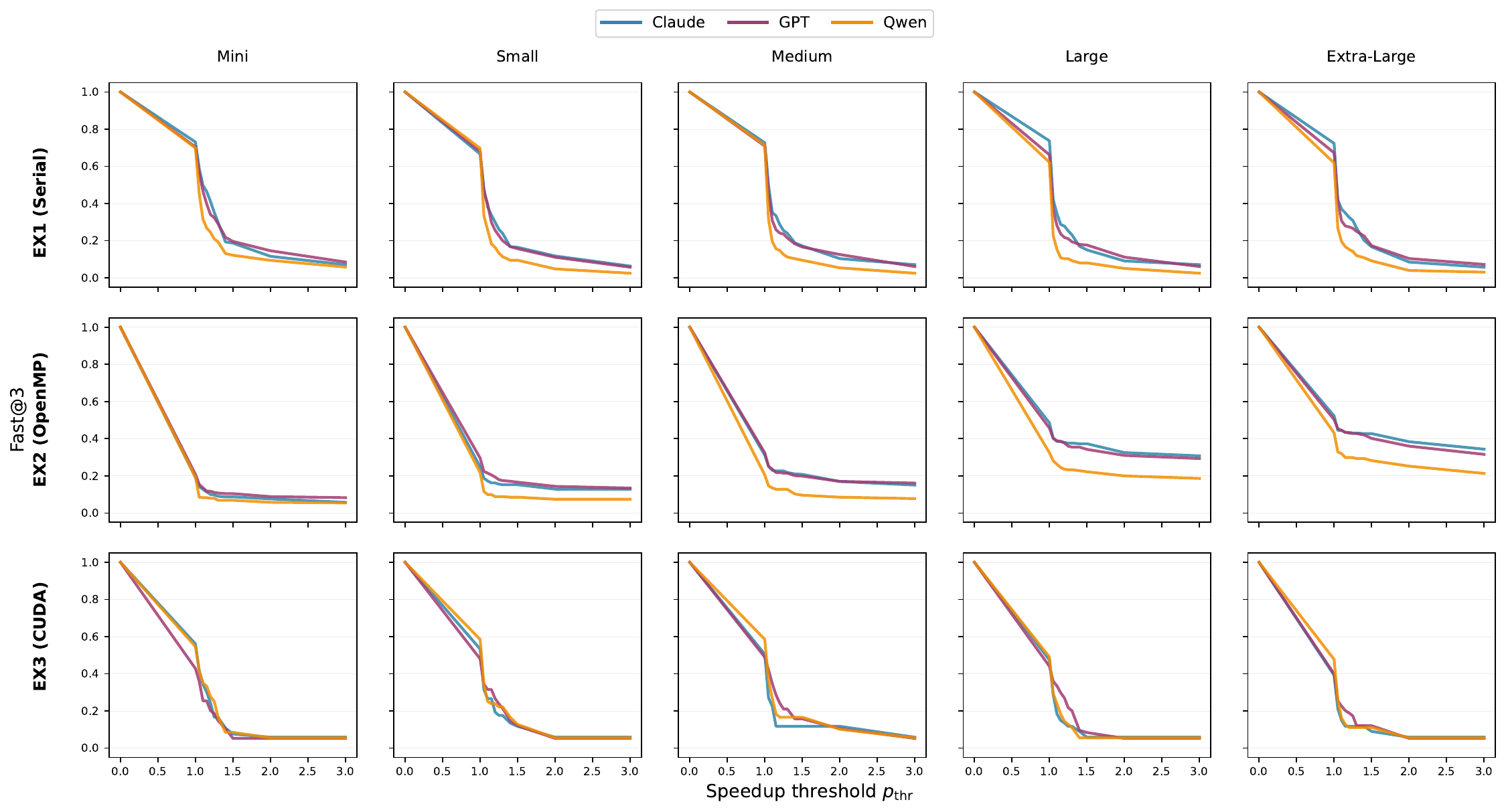}
  \caption{$\mathrm{Fast}@3$ versus $p_{\mathrm{thr}}$ by input size (columns) for EX1, EX2 (16 threads), and EX3 (rows).}
  \label{fig:appendix_fast_per_dataset}
\end{figure*}

\begin{figure*}[t]
  \centering
  \includegraphics[width=0.95\textwidth]{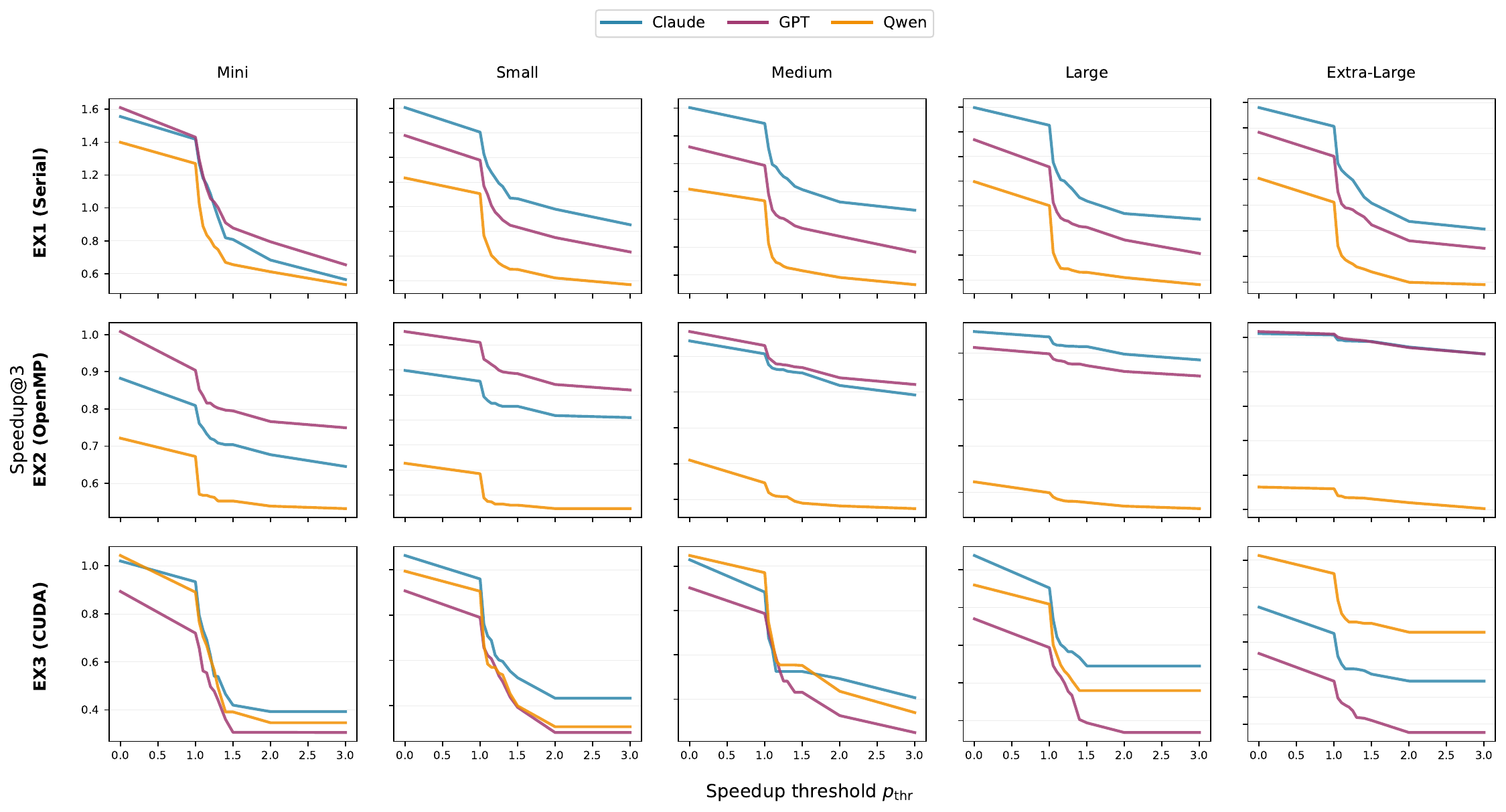}
  \caption{$\mathrm{Speedup}@3$ versus $p_{\mathrm{thr}}$ by input size (columns) for EX1, EX2 (16 threads), and EX3 (rows).}
  \label{fig:appendix_speedup_per_dataset}
\end{figure*}

\Cref{fig:appendix_fast_per_dataset,fig:appendix_speedup_per_dataset} separate the five input sizes aggregated in the main paper \Cref{fig:rq1_rq2}.
Across sizes, $\mathrm{Fast}@3$ falls as $p_{\mathrm{thr}}$ increases, with a sharper drop in EX3.

\subsection{EX1 \& EX2 Speedup@3 Plateau Analysis}
\label{app_subsec:ex1_ex2_speedup_plateau}
\label{app_subsec:ex2_speedup_plateau}
\label{app_subsec:ex2_distribution}

\begin{figure*}[t]
  \centering
  \includegraphics[width=0.92\textwidth]{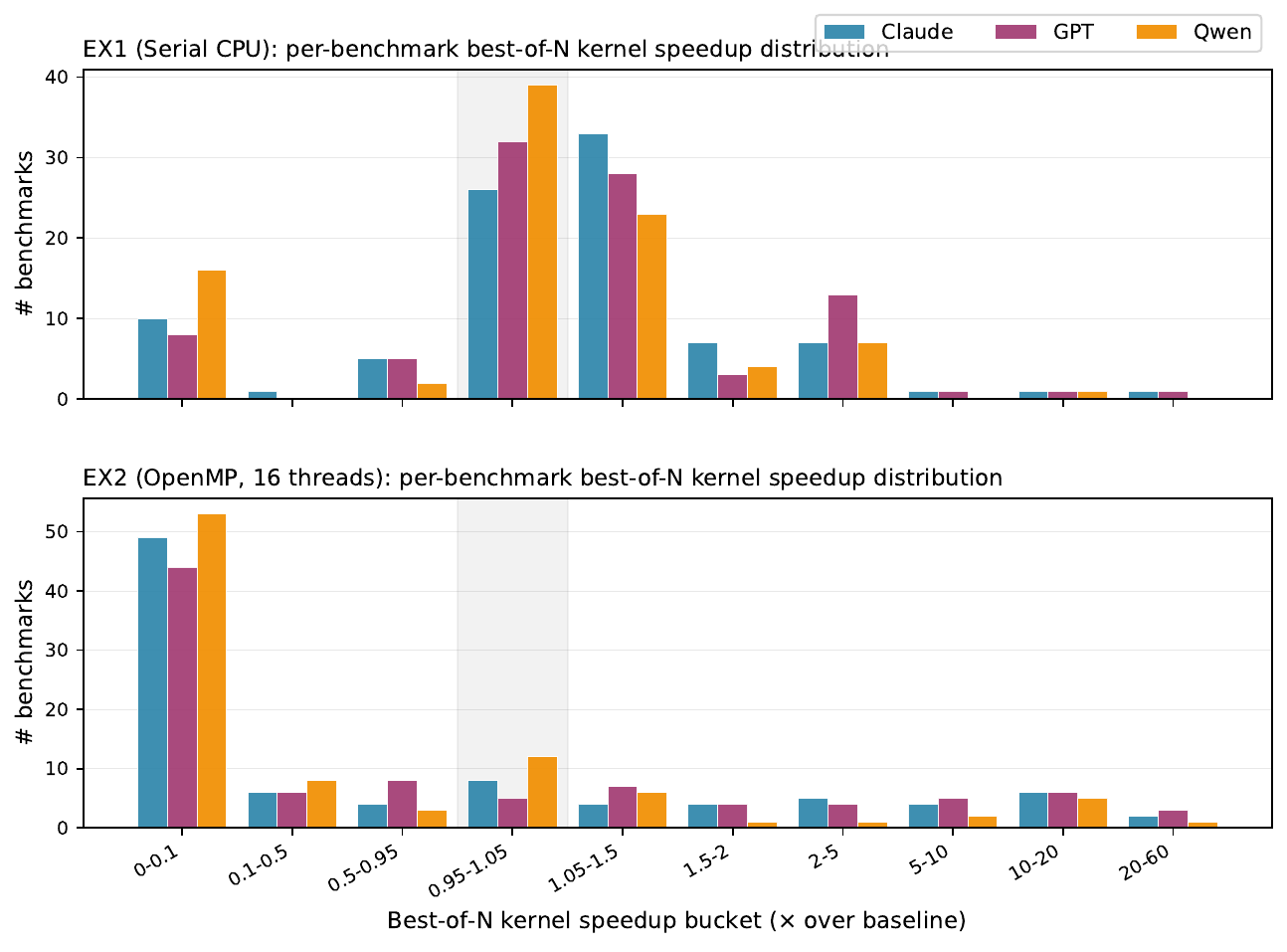}
  \caption{Per-benchmark best-of-$N$ kernel speedup distribution ($N=10$) for EX1 (top) and EX2 at $16$ threads (bottom) on the medium dataset, with $92$ benchmarks per model. Bin widths are non-uniform; the shaded column marks the $[0.95, 1.05{\times}]$ no-gain band.}
  \label{fig:appendix_best_of_n_dist}
\end{figure*}

To examine the EX2 $\mathrm{Speedup}@3$ plateau (\Cref{subsec:rq2_speedup}), we compare per-benchmark best-of-$N$ kernel speedups in EX1 and EX2.
For each model and benchmark on the medium dataset, we take the maximum over $N{=}10$ correctness-gated candidates.
\Cref{fig:appendix_best_of_n_dist} shows the distributions for all three models, and \autoref{tab:bestofn_quantiles} reports summary statistics for Claude.

\textbf{EX1} (top of \autoref{fig:appendix_best_of_n_dist}) is concentrated near baseline: roughly two-thirds of benchmarks fall in $[0.95, 1.5{\times}]$, including $28\%$ in the $[0.95, 1.05{\times}]$ no-gain band.
At most three benchmarks per model exceed $5{\times}$. As $p_{\mathrm{thr}}$ rises, middle-range cases cease to qualify and the macro-average declines.

\textbf{EX2} (bottom) has $44$--$53$ of $92$ benchmarks per model in $[0, 0.1{\times}]$, consistent with high OpenMP overhead relative to short baseline runtimes (\Cref{subsec:motif_difficutly_level}).
The $[0.95, 1.05{\times}]$ no-gain band contains only $5$--$12$ benchmarks (versus $26$--$39$ in EX1), and about $6\%$ fall in $[1.05, 2{\times})$.
At the other extreme, some kernels exceed $20{\times}$, with \texttt{matmul} reaching $48.8{\times}$ for Claude.
These large gains keep the EX2 $\mathrm{Speedup}@3$ macro-average near $2.5$--$2.8{\times}$ over the tested thresholds.

Across broader speedup bins, EX2 places $73\%$ of benchmarks at $\le\!1.05{\times}$, $6\%$ in $[1.05, 2{\times})$, and $18\%$ at $\ge\!2{\times}$.
For EX1, $35\%$ are in $[0.95, 1.05{\times}]$, $30\%$ in $[1.05, 1.5{\times}]$, and approximately $4\%$ reach at least $5{\times}$.

\begin{table}[H]
\centering
\small
\setlength{\tabcolsep}{6pt}
\caption{Summary statistics for Claude's per-benchmark best-of-$N$ kernel speedup on the medium dataset ($N{=}10$, $92$ benchmarks).}
\label{tab:bestofn_quantiles}
\begin{tabular}{lcc}
\toprule
\textbf{Statistic (Claude)} & \textbf{EX1 (Serial)} & \textbf{EX2 (OpenMP, 16t)} \\
\midrule
median                                       & $1.07$ & $0.05$ \\
mean                                         & $1.90$ & $2.43$ \\
std                                          & $4.99$ & $6.45$ \\
$p_{25}$ / $p_{75}$ / $p_{90}$               & $1.00$ / $1.32$ / $2.09$ & $0.00$ / $1.09$ / $8.55$ \\
fraction in $[0.95,1.05{\times}]$ (no-gain)  & $28.3\%$ & $8.7\%$ \\
fraction $\ge 2{\times}$                     & $10.9\%$ & $18.5\%$ \\
fraction in $[0, 0.95{\times}) \cup [2{\times},\infty)$ (extremes) & $28.3\%$ & $82.6\%$ \\
\bottomrule
\end{tabular}
\end{table}

\subsection{Motif- and Structural Complexity-Stratified Results}
\label{app_subsec:stratified_result}
\label{app_subsec:stratfied_result}

\begin{figure*}[t]
  \centering
  \begin{subfigure}{0.49\textwidth}
    \centering
    \includegraphics[width=\textwidth]{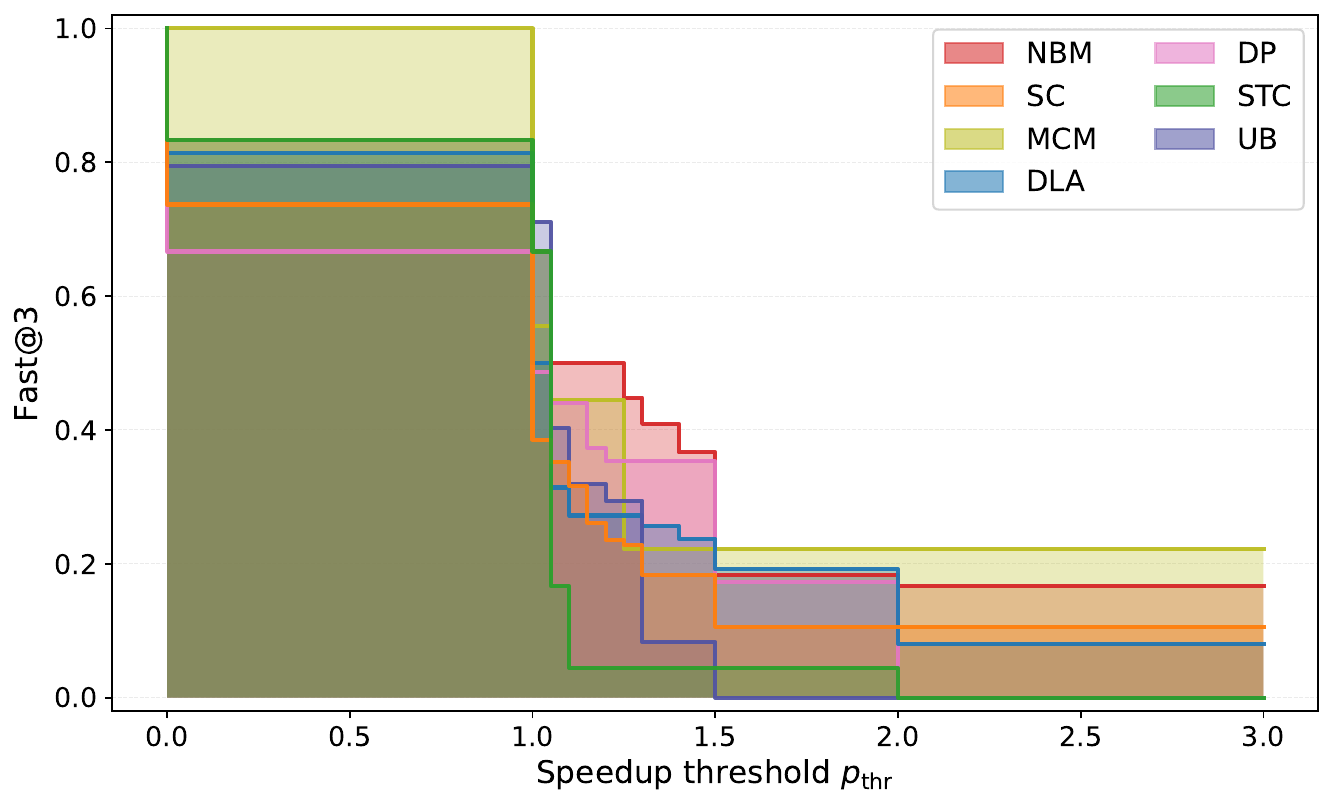}
    \caption{$\mathrm{Fast}@3$, EX1}
    \label{fig:appendix_motif_fast_ex1}
  \end{subfigure}
  \begin{subfigure}{0.49\textwidth}
    \centering
    \includegraphics[width=\textwidth]{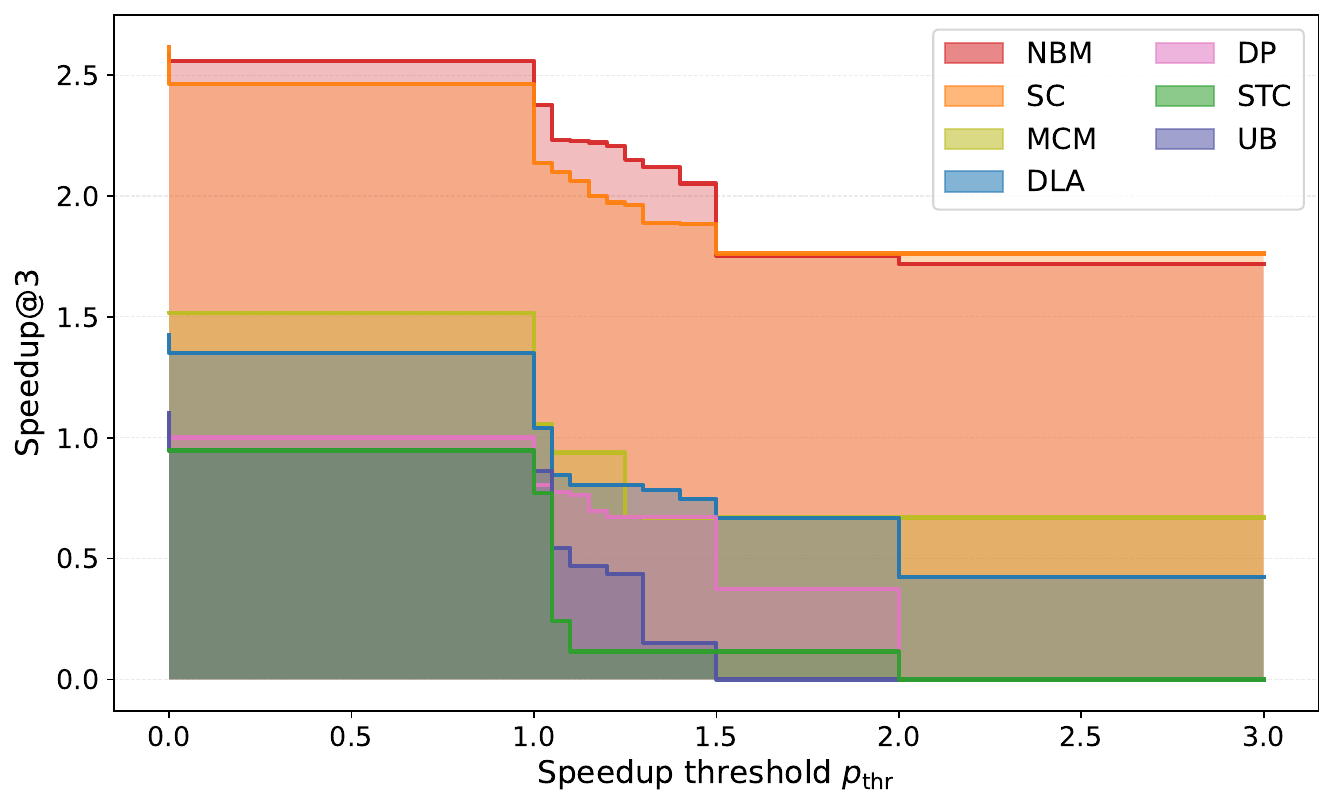}
    \caption{$\mathrm{Speedup}@3$, EX1}
    \label{fig:appendix_motif_speedup_ex1}
  \end{subfigure}
  \begin{subfigure}{0.49\textwidth}
    \centering
    \includegraphics[width=\textwidth]{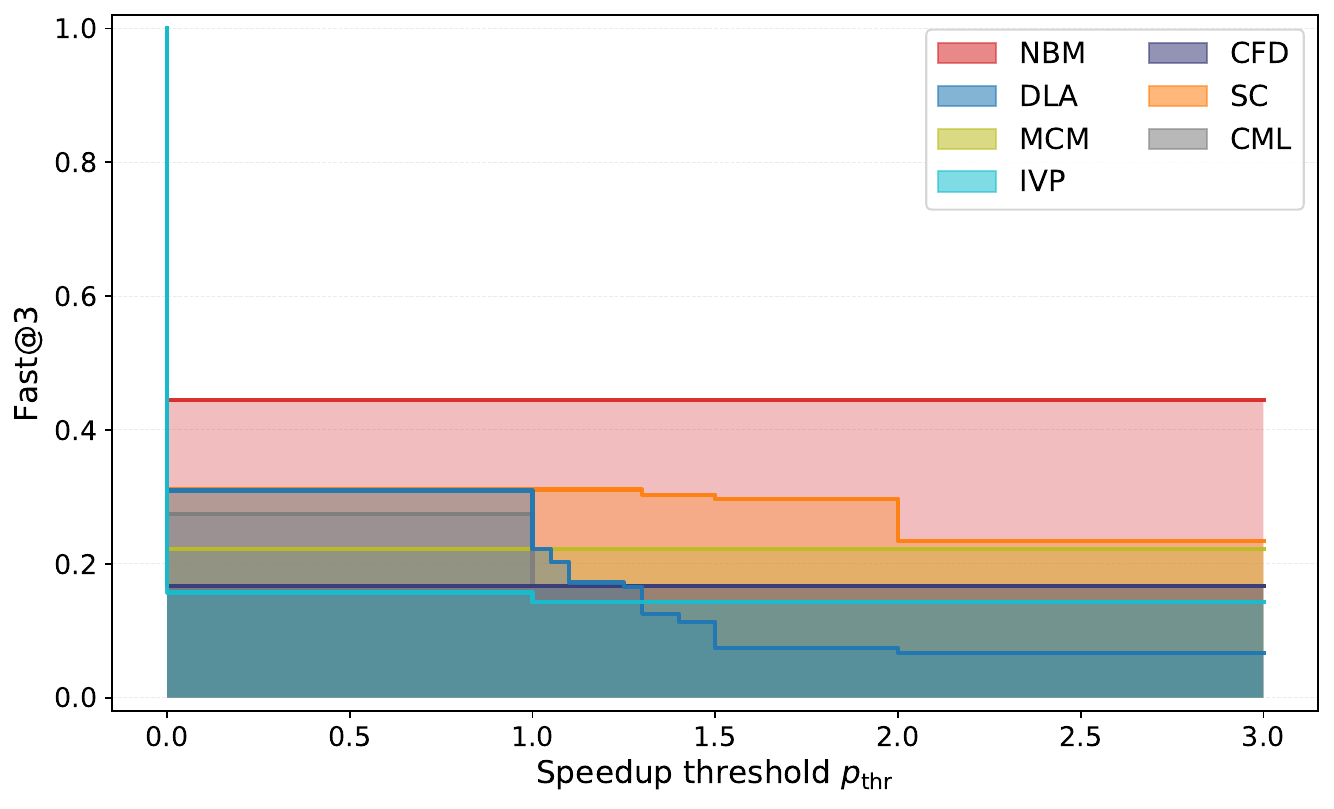}
    \caption{$\mathrm{Fast}@3$, EX2 (16 threads)}
    \label{fig:appendix_motif_fast_ex2}
  \end{subfigure}
  \begin{subfigure}{0.49\textwidth}
    \centering
    \includegraphics[width=\textwidth]{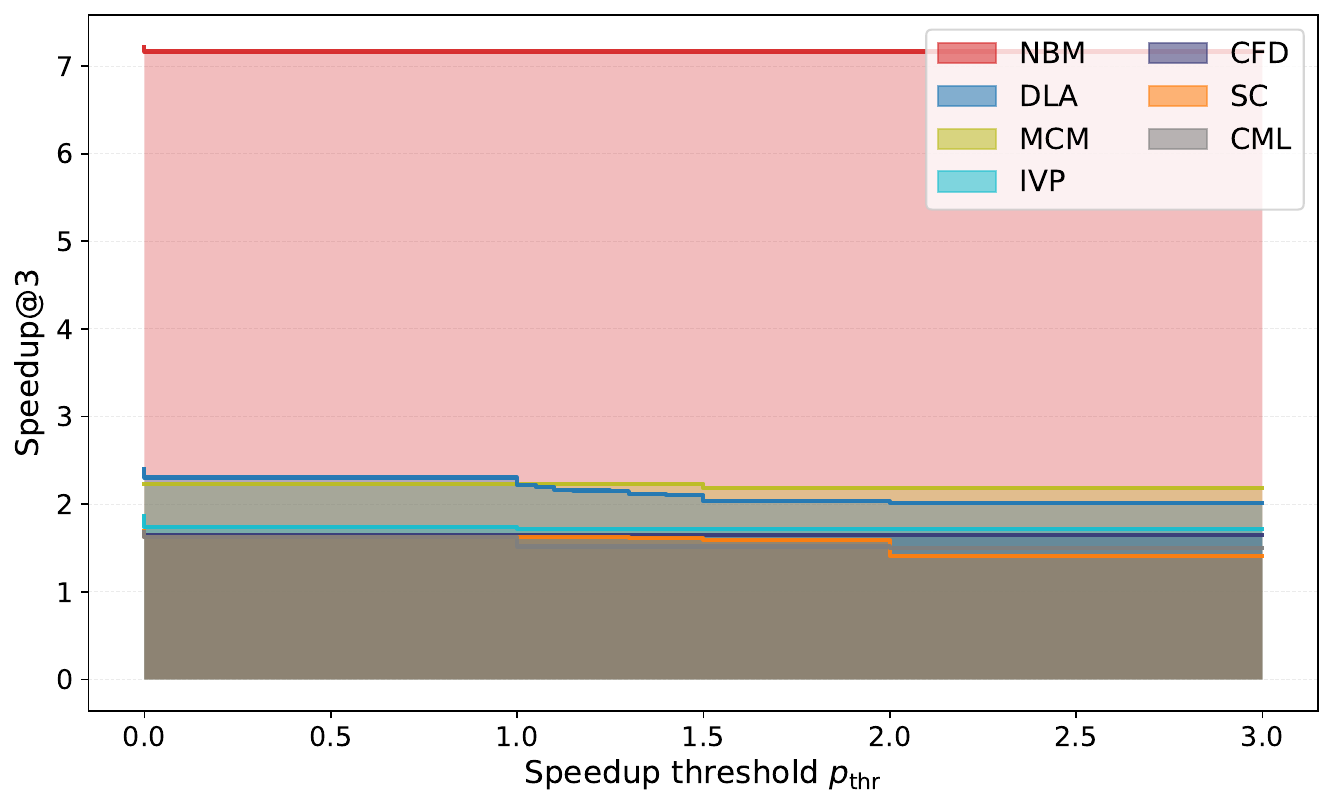}
    \caption{$\mathrm{Speedup}@3$, EX2 (16 threads)}
    \label{fig:appendix_motif_speedup_ex2}
  \end{subfigure}
  \caption{Motif-stratified $\mathrm{Fast}@3$ and $\mathrm{Speedup}@3$ on the medium dataset, macro-averaged across the three models. For legibility, each row shows the top-$7$ motifs by $\mathrm{Speedup}@3$ at $p_{\mathrm{thr}}=1.0$ within that setting (EX1: NBM, SC, MCM, DLA, DP, STC, UB; EX2: NBM, DLA, MCM, IVP, CFD, SC, CML); the remaining motifs are listed in~\Cref{tab:motif_characterization}.}
  \label{fig:appendix_motif_stratified}
\end{figure*}

\begin{figure*}[t]
  \centering
  \begin{subfigure}{0.49\textwidth}
    \centering
    \includegraphics[width=\textwidth]{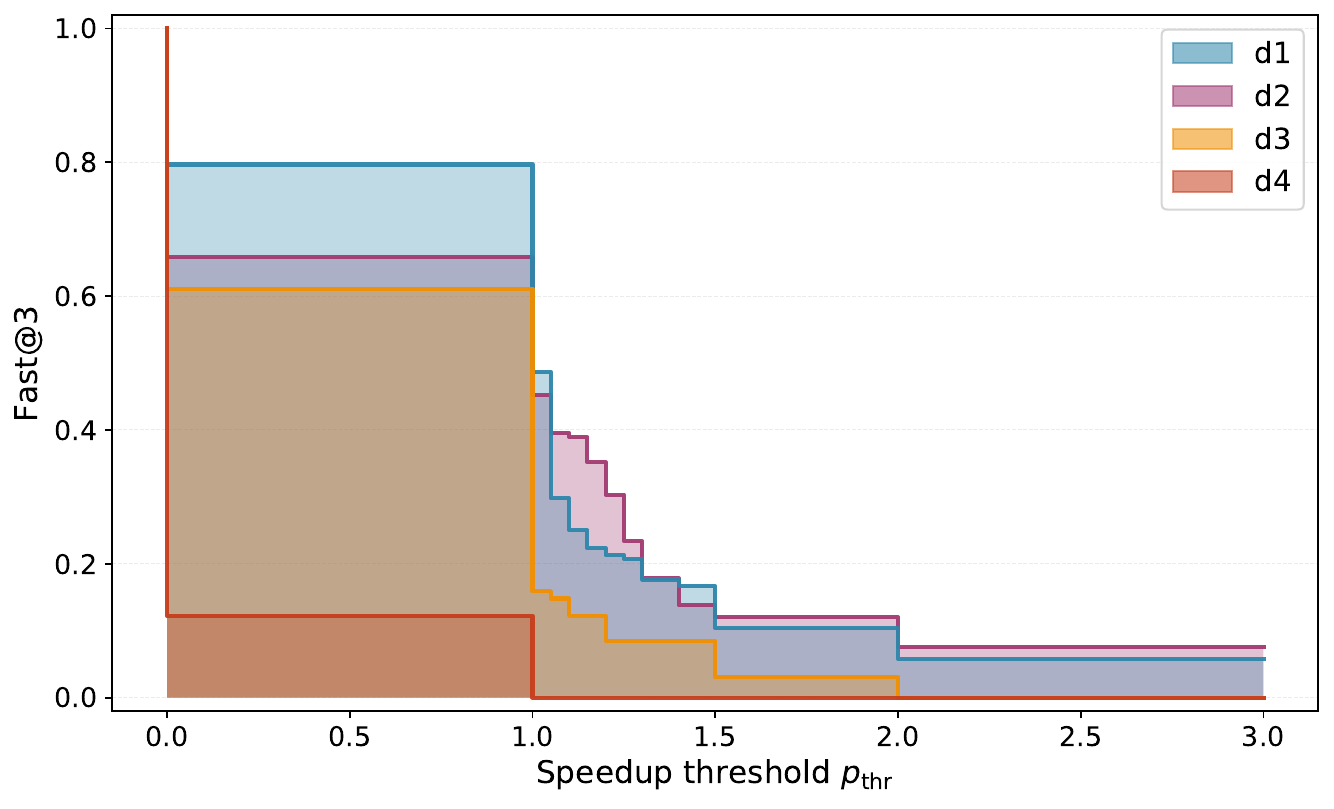}
    \caption{$\mathrm{Fast}@3$, EX1}
    \label{fig:appendix_diff_fast_ex1}
  \end{subfigure}
  \begin{subfigure}{0.49\textwidth}
    \centering
    \includegraphics[width=\textwidth]{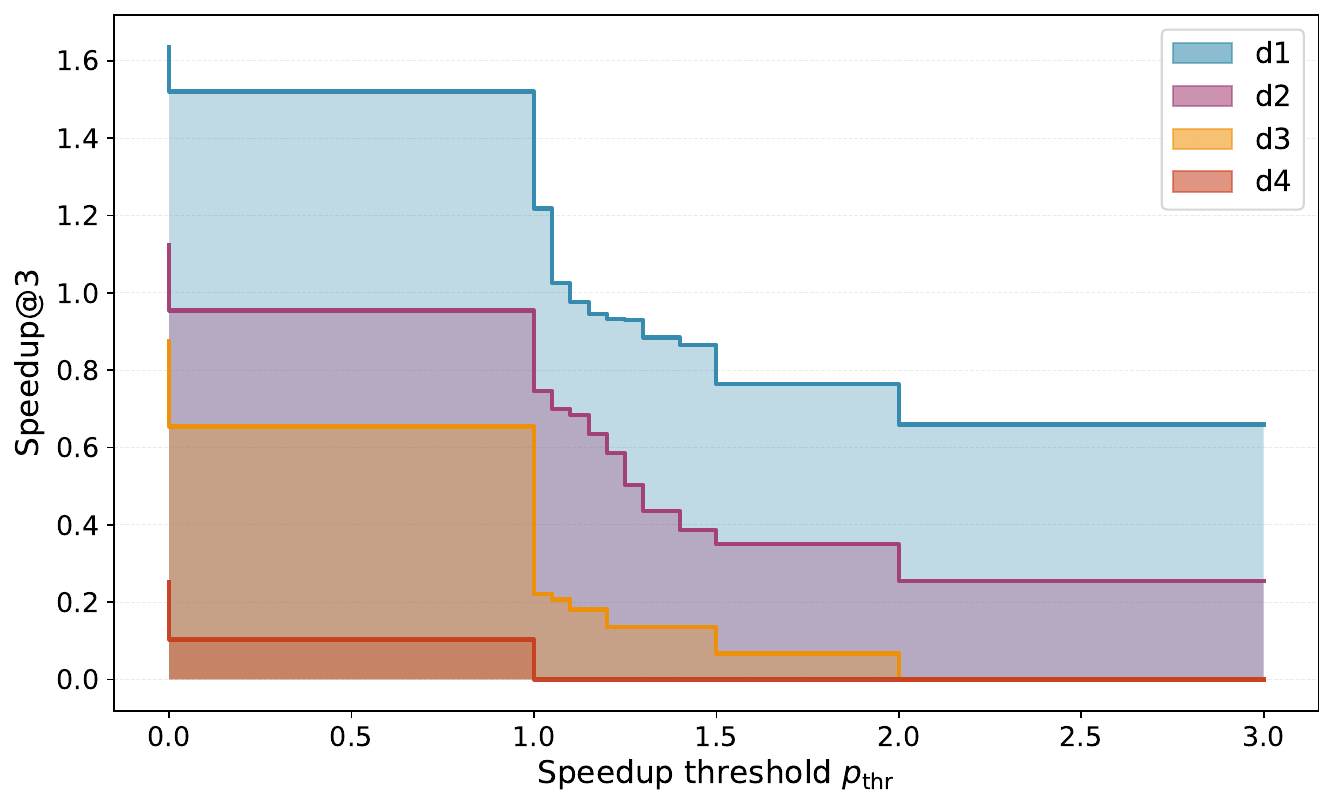}
    \caption{$\mathrm{Speedup}@3$, EX1}
    \label{fig:appendix_diff_speedup_ex1}
  \end{subfigure}
  \begin{subfigure}{0.49\textwidth}
    \centering
    \includegraphics[width=\textwidth]{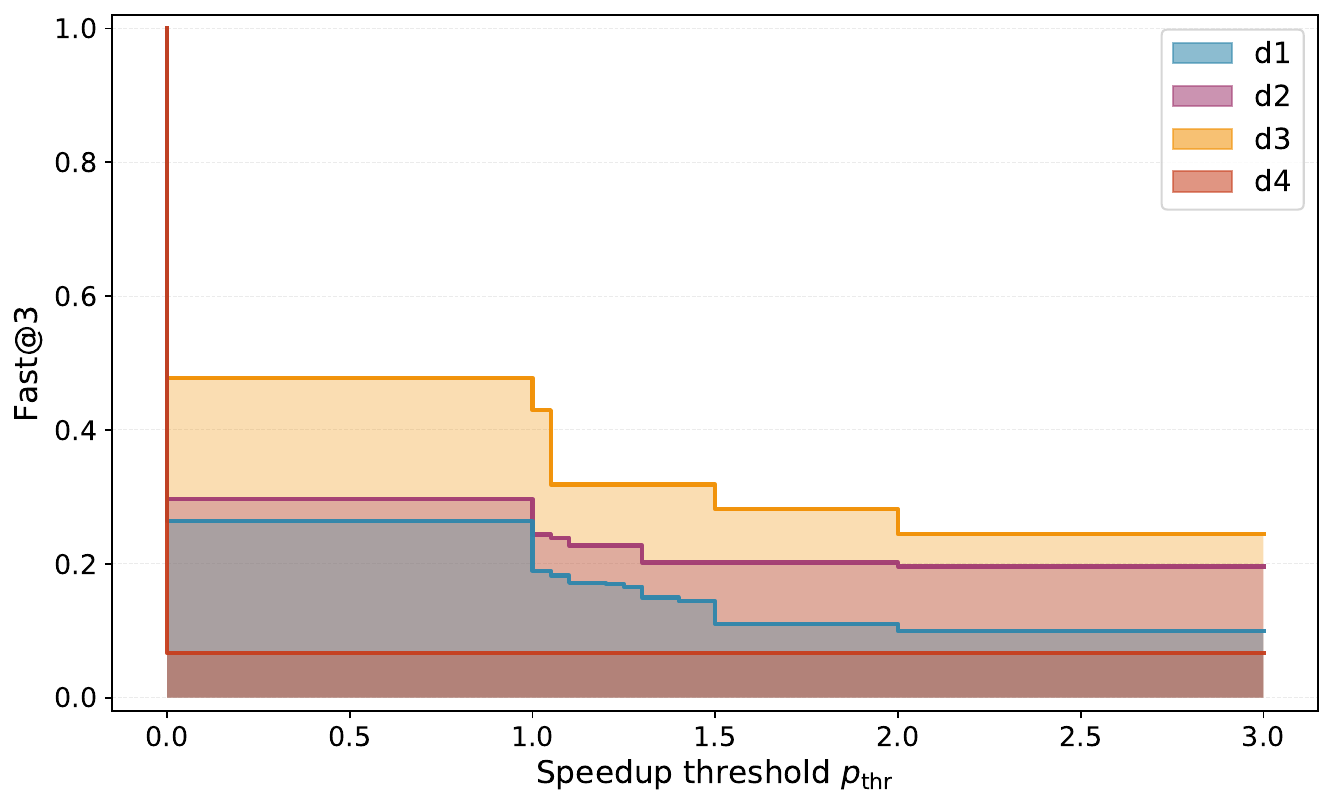}
    \caption{$\mathrm{Fast}@3$, EX2 (16 threads)}
    \label{fig:appendix_diff_fast_ex2}
  \end{subfigure}
  \begin{subfigure}{0.49\textwidth}
    \centering
    \includegraphics[width=\textwidth]{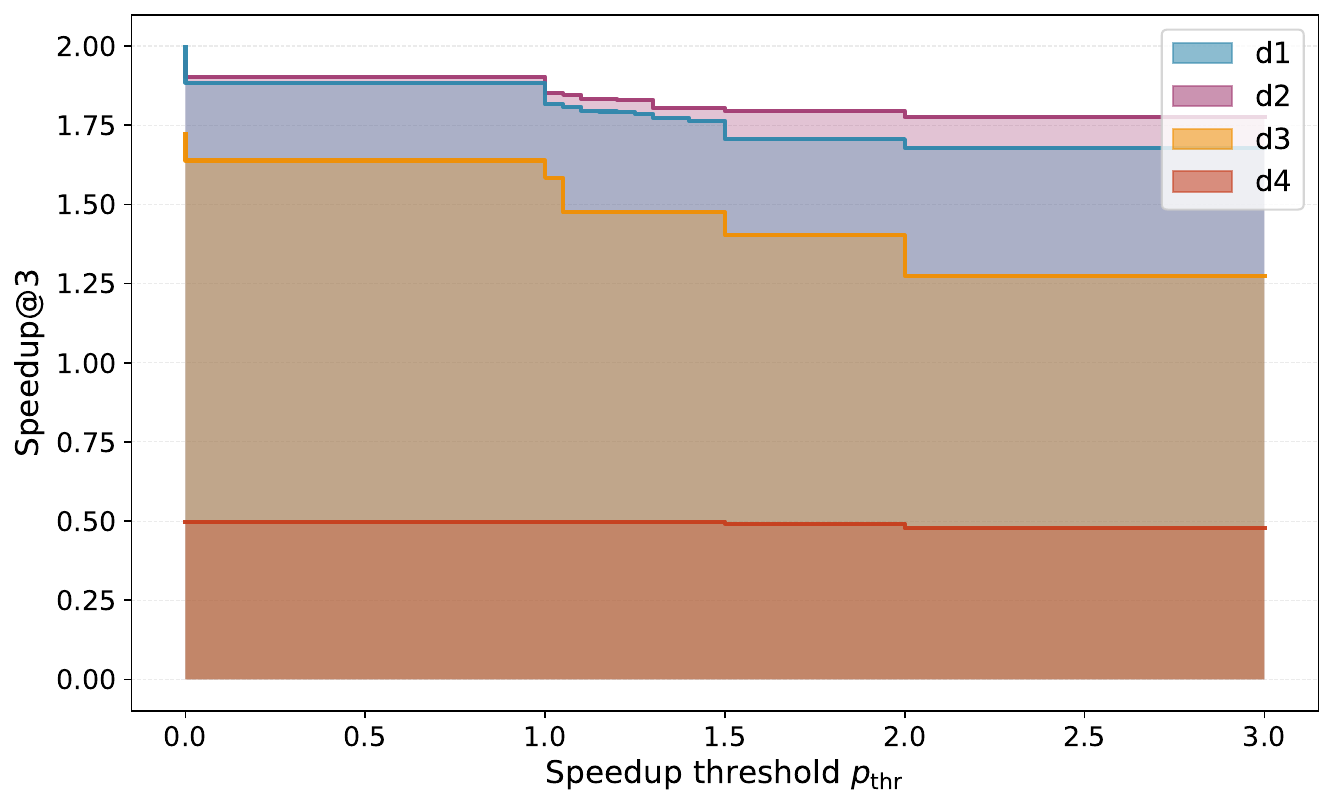}
    \caption{$\mathrm{Speedup}@3$, EX2 (16 threads)}
    \label{fig:appendix_diff_speedup_ex2}
  \end{subfigure}
  \caption{Structural complexity-stratified $\mathrm{Fast}@3$ and $\mathrm{Speedup}@3$ on the medium dataset, macro-averaged across the three models.}
  \label{fig:appendix_difficulty_stratified}
\end{figure*}

\Cref{fig:appendix_motif_stratified} shows motif-stratified $\mathrm{Fast}@3$ and $\mathrm{Speedup}@3$ for EX1 and EX2 on the medium dataset, selecting $7$ motifs with the highest $\mathrm{Speedup}@3$ at $p_{\mathrm{thr}}=1.0$ in each setting (EX1: NBM, SC, MCM, DLA, DP, STC, UB; EX2: NBM, DLA, MCM, IVP, CFD, SC, CML).
The remaining $9$ motifs are listed in~\Cref{tab:motif_characterization} but omitted from the figure for clarity.
\Cref{fig:appendix_difficulty_stratified} shows the corresponding results across all four structural complexity strata ($d_1$--$d_4$).

EX1 generally follows the structural complexity trend, with lower structural complexity strata achieving higher $\mathrm{Fast}@3$ and $\mathrm{Speedup}@3$.
In EX2, short-running $d_1$ kernels are more affected by OpenMP overhead, while longer $d_2$/$d_3$ kernels benefit from parallelization, as discussed in~\Cref{subsec:motif_difficutly_level}.
\Cref{tab:difficulty_ex2} provides the per structural complexity stratum aggregate results, and \Cref{tab:d4_breakdown} gives the per-benchmark $d_4$ breakdown.

\paragraph{Per-Structural-Complexity-Stratum Aggregate.}
\Cref{tab:difficulty_ex2} reports baseline kernel runtime, the count below $1.05\times$ Best-of-$N$ speedup at $16$ threads, and the largest speedups for each structural complexity stratum.
All $92$ CPU/OpenMP benchmarks have valid single-thread baselines on the medium dataset.
The $d_1$ group has the largest fraction below $1.05\times$ ($67\%$), but also includes \texttt{matmul} ($49.6{\times}$), \texttt{haccmk} ($30.7{\times}$), and \texttt{lavaMD} ($21.5{\times}$).
These outcomes reflect differences in runtime scale and available parallelism within the same structural complexity stratum.

\begin{table}[H]
\centering
\small
\setlength{\tabcolsep}{4pt}
\caption{Per-structural complexity stratum EX2 outcomes at $16$ threads on the medium dataset, restricted to benchmarks with valid EX2 single-thread baselines. ``Fail $<\!1.05{\times}$'' counts benchmarks whose best-of-$N$ kernel speedup over all $30$ trials ($N{=}10$ per model) does not reach $1.05{\times}$.}
\label{tab:difficulty_ex2}
\begin{tabularx}{\textwidth}{l c c c c X}
\toprule
\textbf{Structural Complexity} & \textbf{\#B} & \textbf{Median 1t baseline} & \textbf{Fail $<\!1.05{\times}$} & \textbf{Top speedup} & \textbf{Top-3 kernels (best-of-$N$)} \\
\midrule
$d_1$ & $61$ & $7.33$\,ms & $41$ ($67\%$) & $49.63{\times}$ & \texttt{matmul} $49.6{\times}$, \texttt{haccmk} $30.7{\times}$, \texttt{lavaMD} $21.5{\times}$ \\
$d_2$ & $17$ & $1.05$\,s  & $5$  ($29\%$) & $13.27{\times}$ & \texttt{hotspot3D} $13.3{\times}$, \texttt{kmeans} $12.8{\times}$, \texttt{particlefilter} $12.4{\times}$ \\
$d_3$ & $9$  & $1.37$\,s  & $1$  ($11\%$) & $8.49{\times}$  & \texttt{b+tree} $8.5{\times}$, \texttt{nw} $7.4{\times}$, \texttt{hotspot} $5.3{\times}$ \\
$d_4$ & $5$  & $6.27$\,s  & $2$  ($40\%$) & $8.79{\times}$  & \texttt{heartwall} $8.8{\times}$, \texttt{susan-e-mibench} $8.3{\times}$, \texttt{susan-c-mibench} $7.3{\times}$ \\
\bottomrule
\end{tabularx}
\end{table}

\paragraph{Per-benchmark $d_4$ Breakdown.}
\label{app_subsec:d4_breakdown}
\Cref{tab:d4_breakdown} reports EX2 results for the $5$ $d_4$ benchmarks.

\begin{table}[H]
\centering
\small
\setlength{\tabcolsep}{4pt}
\caption{$d_4$ benchmark outcomes under EX2 at $16$ threads on the medium dataset. Each benchmark has $30$ candidate trials ($N{=}10$ per model). NC = no successful candidate output (compilation or runtime failure); CE-incorrect = produced output but failed correctness check; correct = passed correctness check. ``Best correct $\mathrm{spd}$'' is the maximum kernel speedup over all correct candidates.}
\label{tab:d4_breakdown}
\begin{tabular}{lcccccc}
\toprule
\textbf{Benchmark} & \textbf{Motif} & \textbf{1t baseline} & \textbf{NC} & \textbf{CE-incorrect} & \textbf{Correct} & \textbf{Best correct spd} \\
\midrule
\texttt{cutcp}            & NBM & $6.27$\,s    & $22/30$ & $8/30$  & $0/30$  & --- \\
\texttt{lbm}              & CFD & $9.26$\,s    & $21/30$ & $9/30$  & $0/30$  & --- \\
\texttt{heartwall}        & MSC & $21.73$\,s   & $24/30$ & $0/30$  & $6/30$  & $8.79{\times}$ (GPT-5.1) \\
\texttt{susan-c-mibench}  & IVP & $590.60$\,ms & $19/30$ & $2/30$  & $9/30$  & $7.33{\times}$ (Claude) \\
\texttt{susan-e-mibench}  & IVP & $939.61$\,ms & $12/30$ & $0/30$  & $18/30$ & $8.34{\times}$ (Claude) \\
\bottomrule
\end{tabular}
\end{table}

\texttt{cutcp} and \texttt{lbm} produce no correct candidates, while only GPT-5.1 produces correct \texttt{heartwall} candidates ($6/10$, mean kernel speedup $6.43{\times}$, range from $1.90$ to $8.79{\times}$);
Claude and Qwen each fail to compile all $10$ candidates.
The two MiBench \texttt{susan} variants achieve correct fractions of $9/30$ and $18/30$, with best speedups of $7.33{\times}$ and $8.34{\times}$. 
Thus, high structural complexity does not always imply low correctness.

\subsection{Sampling-Budget Sensitivity for EX2 and EX3}
\label{app_subsec:sampling_budget_sensitivity}

\begin{figure*}[t]
  \centering
  \begin{subfigure}{0.49\textwidth}
    \centering
    \includegraphics[width=\textwidth]{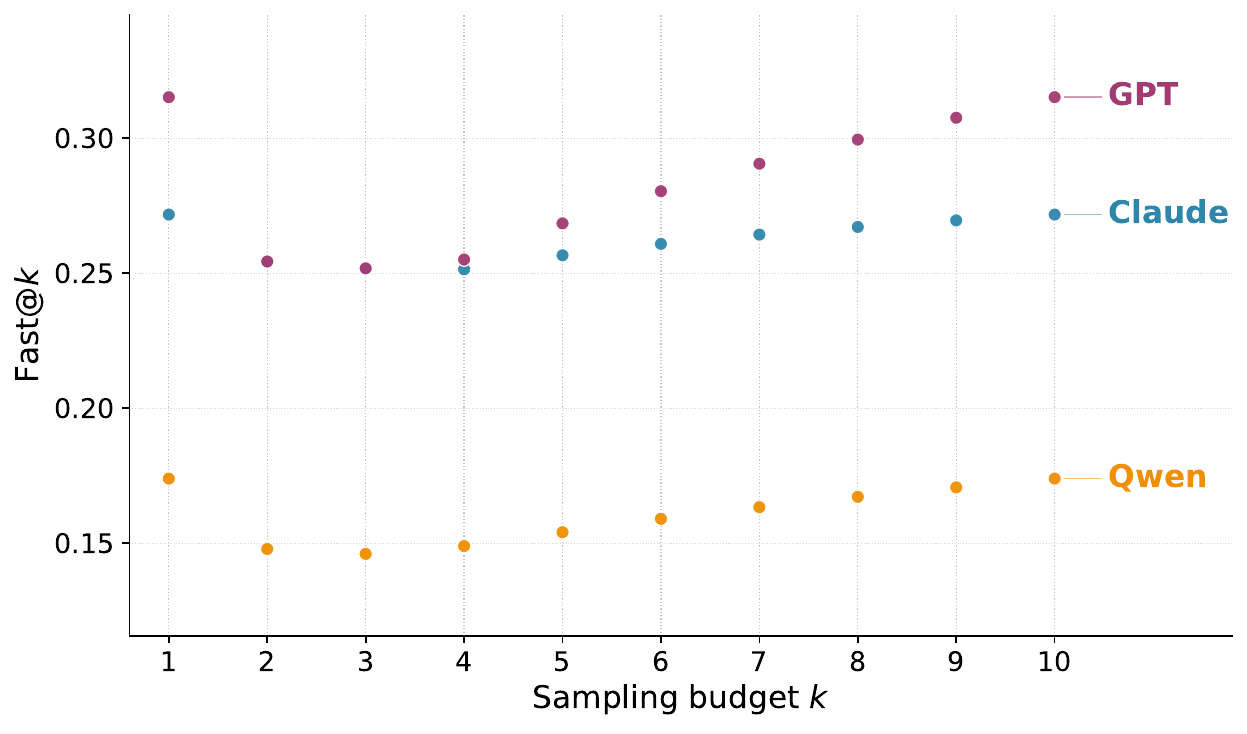}
    \caption{$\mathrm{Fast}@k$ vs.\ $k$, EX2 (16 threads)}
    \label{fig:appendix_fast_k_ex2}
  \end{subfigure}
  \begin{subfigure}{0.49\textwidth}
    \centering
    \includegraphics[width=\textwidth]{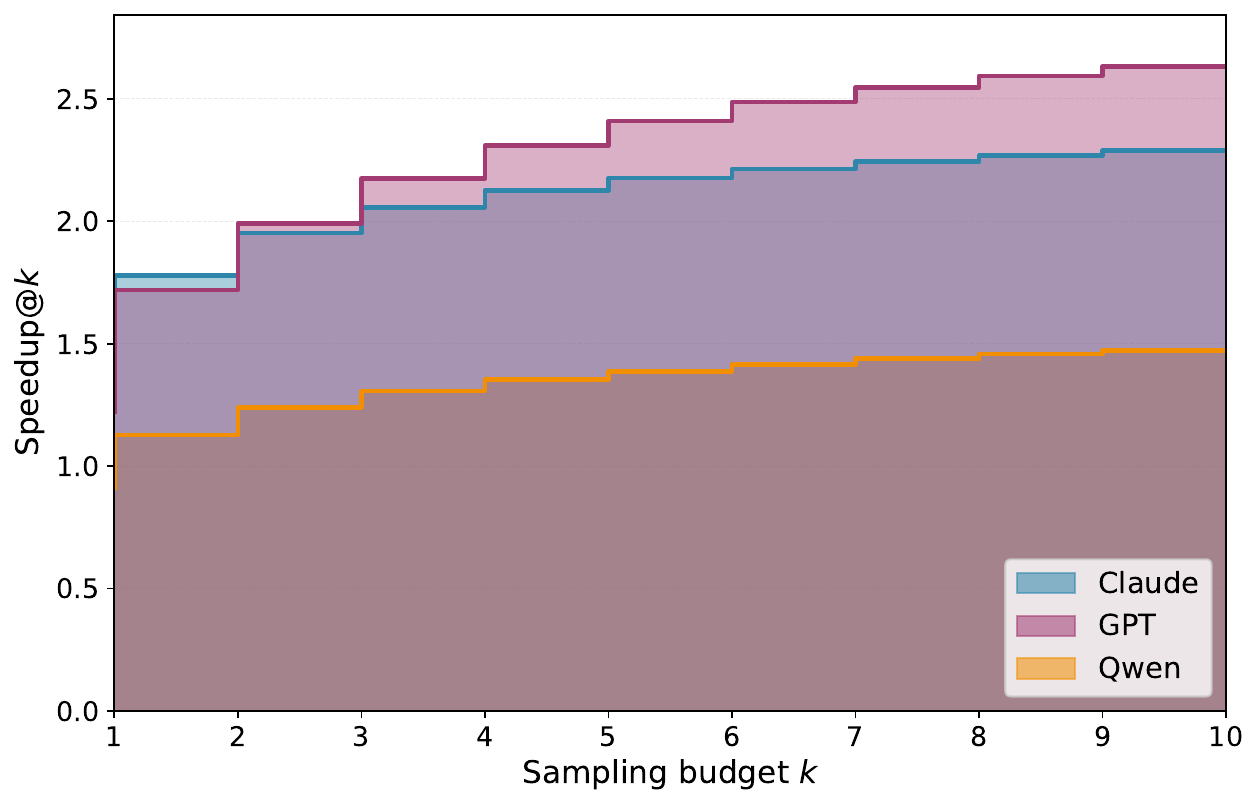}
    \caption{$\mathrm{Speedup}@k$ vs.\ $k$, EX2 (16 threads)}
    \label{fig:appendix_speedup_k_ex2}
  \end{subfigure}
  \caption{EX2 sampling-budget sensitivity at $p_{\mathrm{thr}}=1.05$ on the medium dataset.}
  \label{fig:appendix_k_sensitivity_ex2}
\end{figure*}

\begin{figure*}[t]
  \centering
  \begin{subfigure}{0.49\textwidth}
    \centering
    \includegraphics[width=\textwidth]{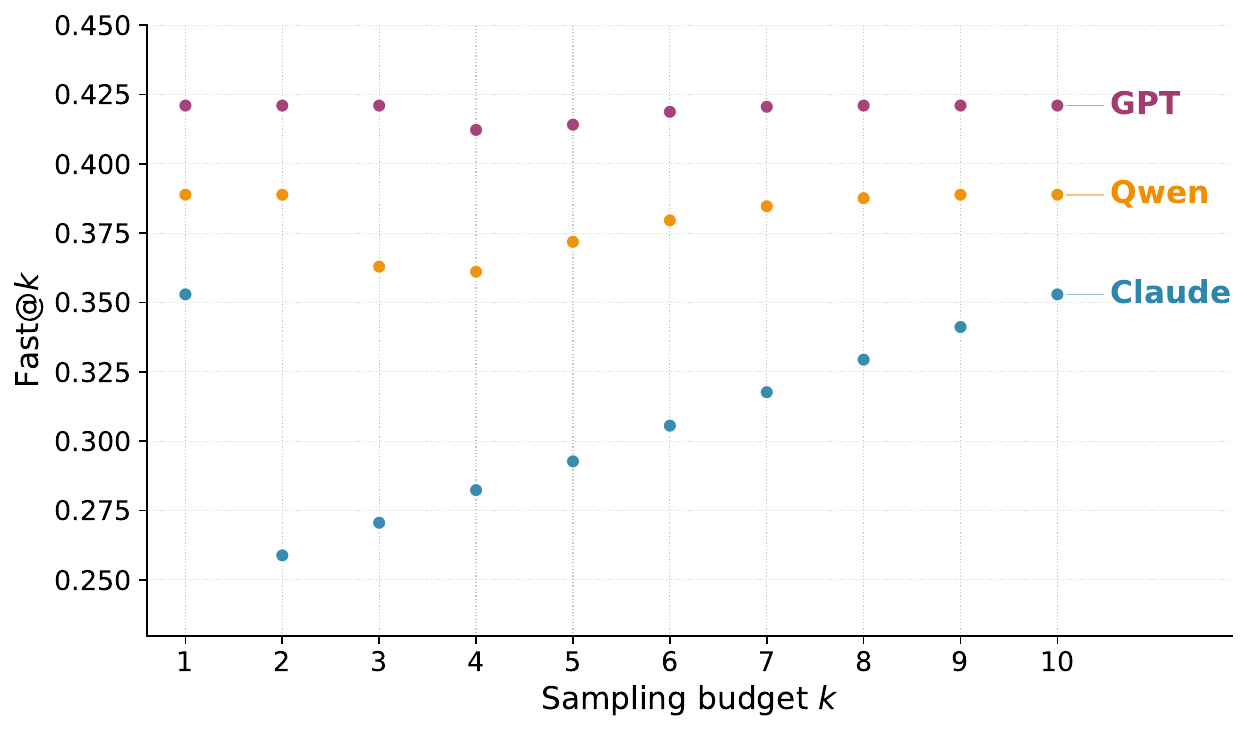}
    \caption{$\mathrm{Fast}@k$ vs.\ $k$, EX3}
    \label{fig:appendix_fast_k_ex3}
  \end{subfigure}
  \begin{subfigure}{0.49\textwidth}
    \centering
    \includegraphics[width=\textwidth]{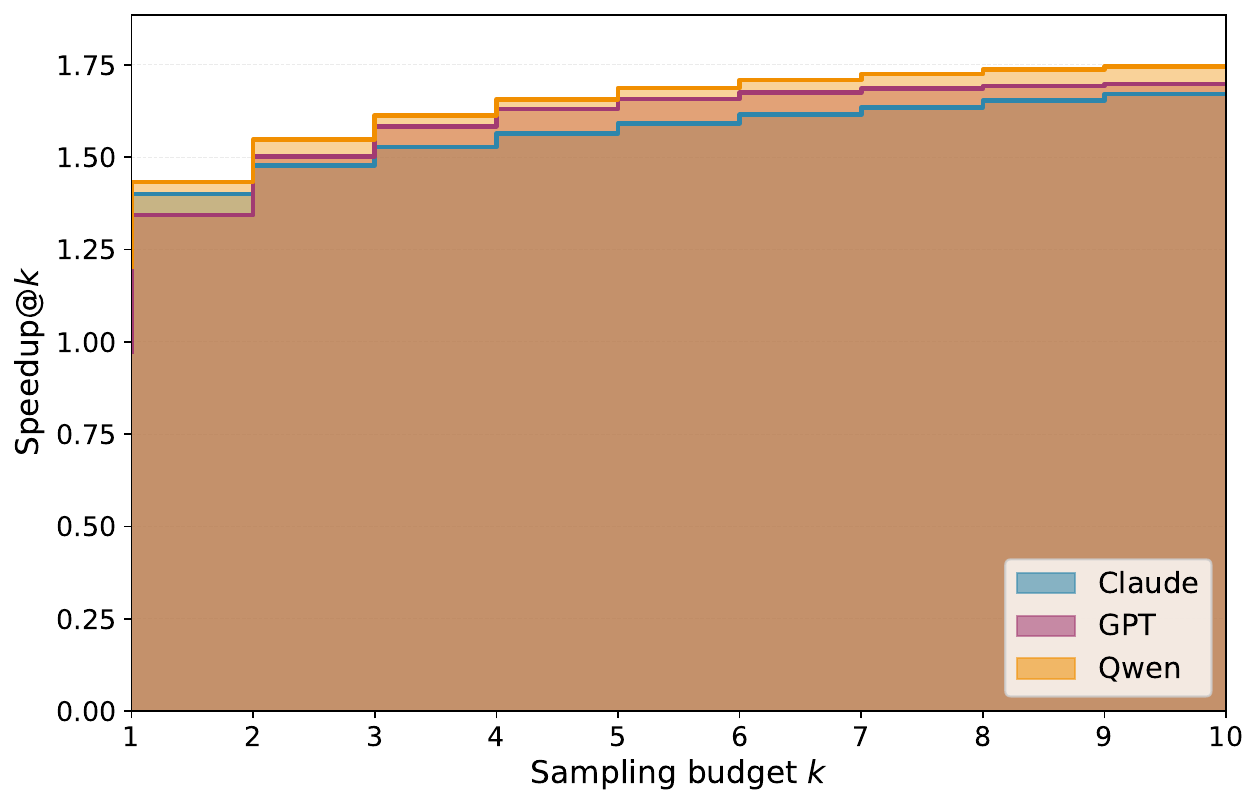}
    \caption{$\mathrm{Speedup}@k$ vs.\ $k$, EX3}
    \label{fig:appendix_speedup_k_ex3}
  \end{subfigure}
  \caption{EX3 sampling-budget sensitivity at $p_{\mathrm{thr}}=1.05$ on the medium dataset.}
  \label{fig:appendix_k_sensitivity_ex3}
\end{figure*}

\Cref{subsubsec:sampling_budget_sensitivity} presents the sampling-budget sensitivity results for EX1, while \Cref{fig:appendix_k_sensitivity_ex2,fig:appendix_k_sensitivity_ex3} show the corresponding $\mathrm{Fast}@k$ and $\mathrm{Speedup}@k$ curves for EX2 ($16$ threads) and EX3 on the medium dataset, with $p_{\mathrm{thr}}=1.05$ and $k\in\{1,\dots,10\}$.

In EX2, $\mathrm{Fast}@k$ at $p_{\mathrm{thr}}=1.05$ is roughly stable as $k$ increases, while $\mathrm{Speedup}@k$ grows noticeably with $k$, dominated by the same high-parallel-speedup kernels identified in~\Cref{app_subsec:ex2_speedup_plateau}.
In EX3, both metrics change modestly with $k$, and the three models remain close.

\subsection{Thread-Count Sensitivity for EX2}
\label{app_subsec:thread_sensitivity}

\begin{figure*}[t]
  \centering
  \begin{subfigure}{0.49\textwidth}
    \centering
    \includegraphics[width=\textwidth]{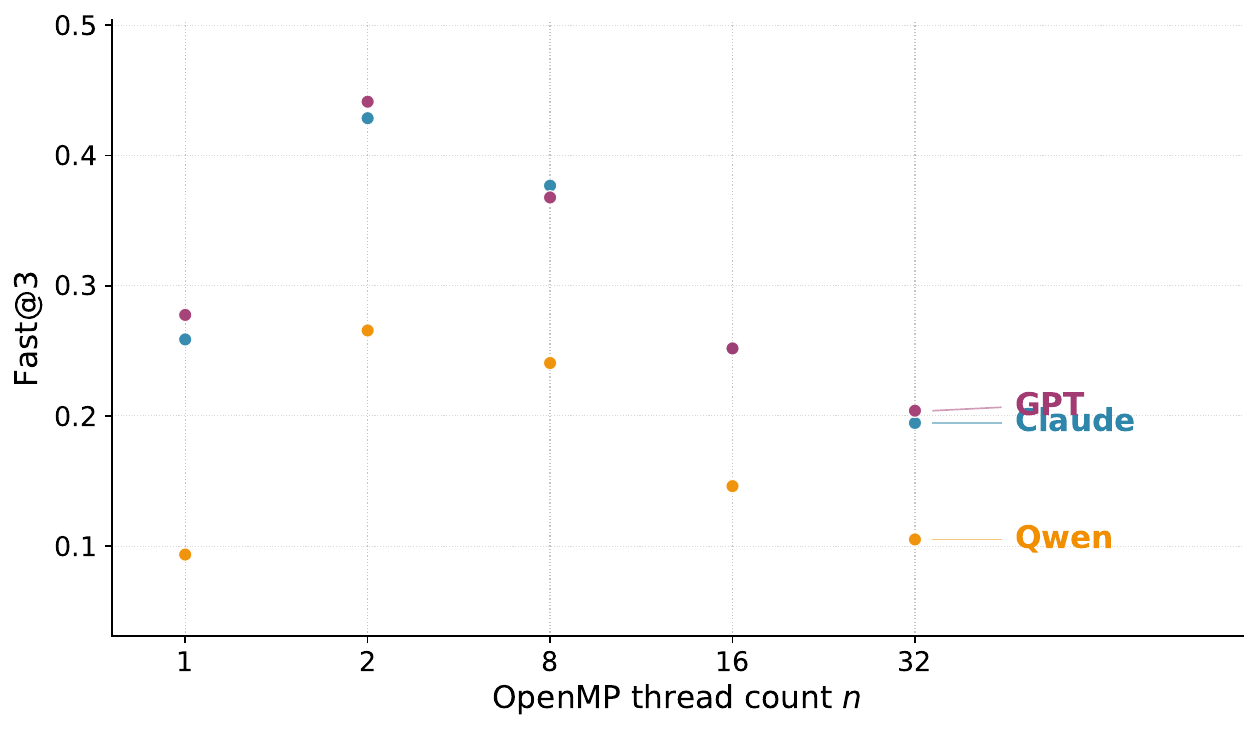}
    \caption{$\mathrm{Fast}@3$ vs.\ $n$, EX2}
    \label{fig:appendix_fast_threads}
  \end{subfigure}
  \begin{subfigure}{0.49\textwidth}
    \centering
    \includegraphics[width=\textwidth]{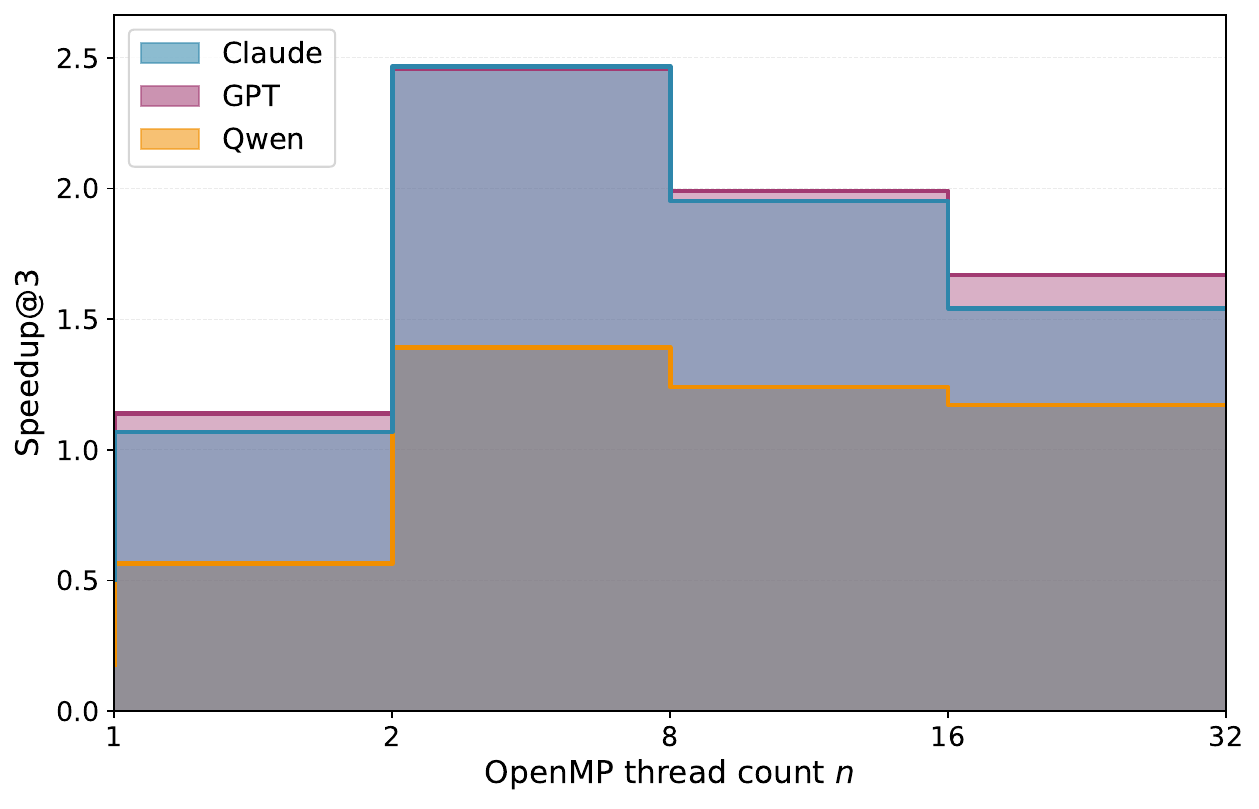}
    \caption{$\mathrm{Speedup}@3$ vs.\ $n$, EX2}
    \label{fig:appendix_speedup_threads}
  \end{subfigure}
  \caption{EX2 thread-count sensitivity at $p_{\mathrm{thr}}=1.05$ on the medium dataset, with $n\in\{1,2,8,16,32\}$.}
  \label{fig:appendix_thread_sensitivity}
\end{figure*}

The main paper reports EX2 results at $16$ threads, while \Cref{fig:appendix_thread_sensitivity} compares EX2 with $n\in\{1,2,8,16,32\}$ threads on the medium dataset at $p_{\mathrm{thr}}=1.05$.
$\mathrm{Speedup}@3$ peaks around $n=8$ and decreases beyond $n=16$, consistent with increasing core contention and OpenMP overhead for moderate-runtime kernels.
$\mathrm{Fast}@3$ peaks at $n=2$ and declines at larger thread counts, as parallelization overhead prevents more short-running kernels from reaching the speedup threshold.
Both decline at higher thread counts, consistent with OpenMP overhead affecting short-running kernels.
At $n=1$, the OpenMP runtime remains active.

\subsection{Total Runtime}
\label{app_subsec:total_runtime_metrics}

\begin{figure*}[t]
  \centering
  \includegraphics[width=0.95\textwidth]{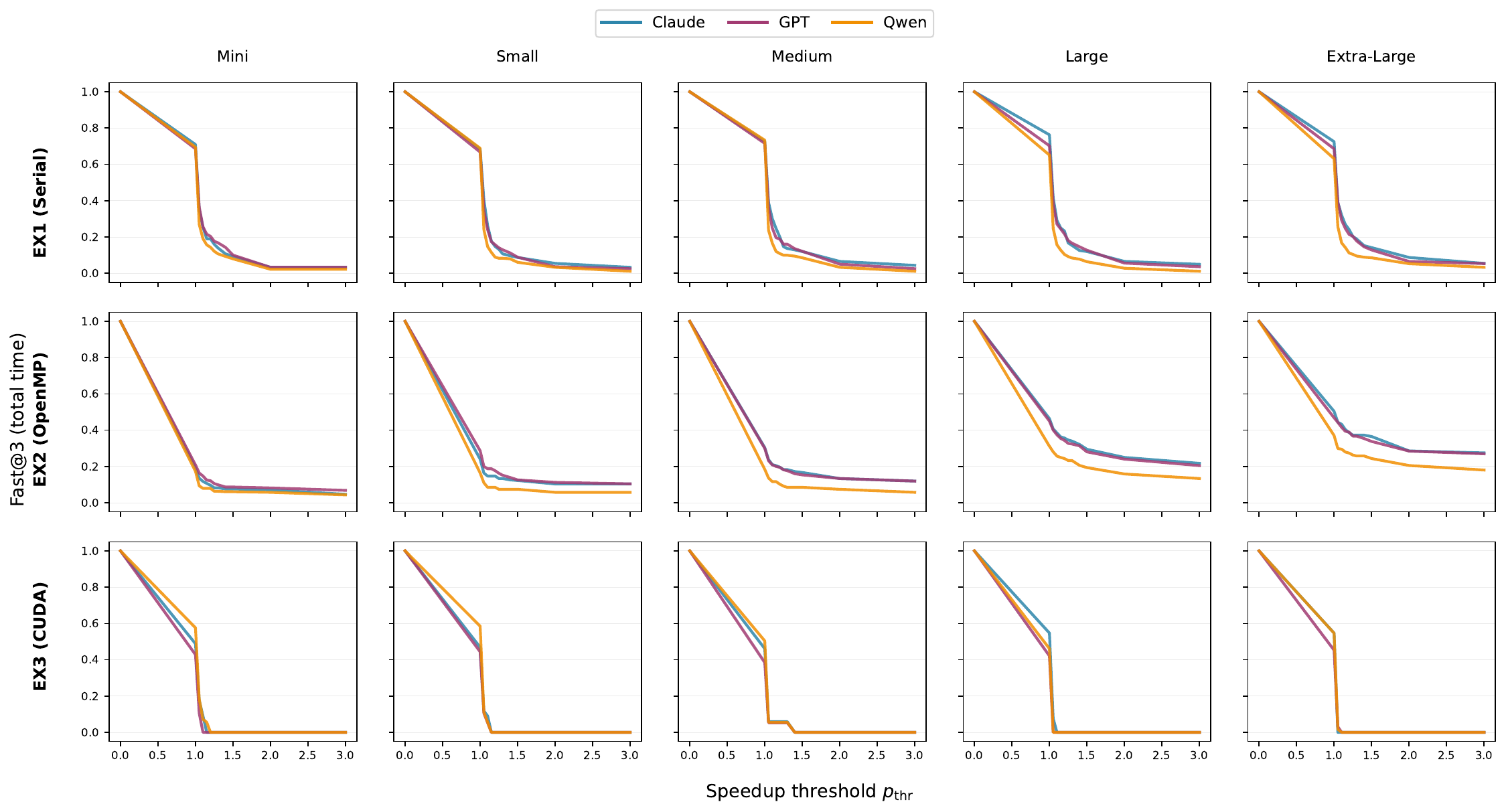}
  \caption{$\mathrm{Fast}@3$ versus $p_{\mathrm{thr}}$ using total program runtime for EX1--EX3 across all input sizes (EX2: $16$ threads).}
  \label{fig:appendix_fast_total}
\end{figure*}

\begin{figure*}[t]
  \centering
  \includegraphics[width=0.95\textwidth]{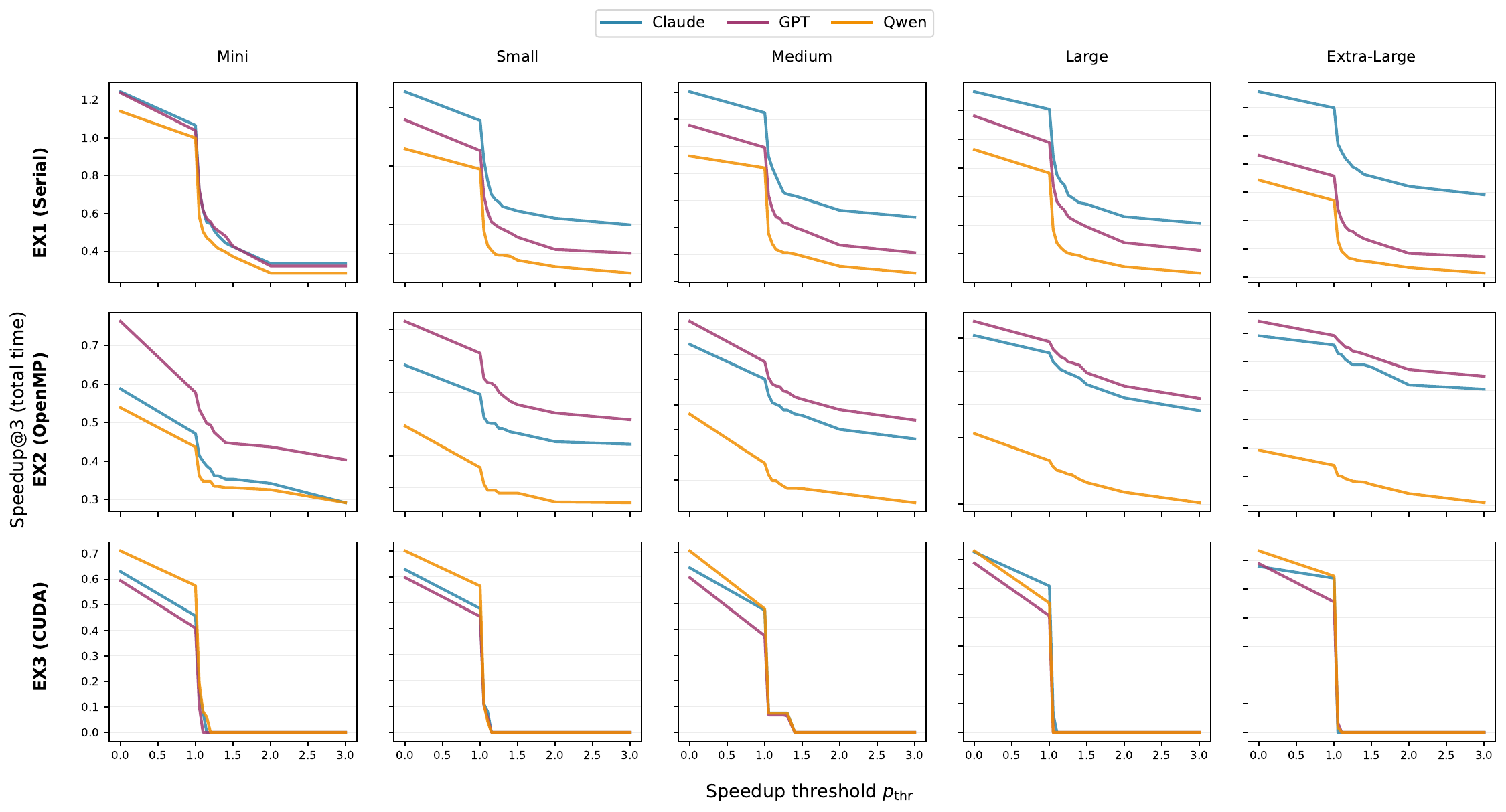}
  \caption{$\mathrm{Speedup}@3$ versus $p_{\mathrm{thr}}$ using total program runtime for EX1--EX3 across all input sizes (EX2: $16$ threads).}
  \label{fig:appendix_speedup_total}
\end{figure*}

The main paper reports kernel-time speedups because \HPCBench optimizes the annotated kernel regions.
\Cref{fig:appendix_fast_total,fig:appendix_speedup_total} report the corresponding $\mathrm{Fast}@3$ and $\mathrm{Speedup}@3$ results using total program runtime.
Total-runtime speedups are generally lower because they include input setup, allocation, I/O, and, for EX3, PCIe transfers and CUDA context initialization, which occur outside the LLM-edited regions.
The qualitative trends remain similar, including the distinction between optimization reliability and achieved speedup.
For EX3, total-runtime $\mathrm{Speedup}@3$ falls to nearly zero beyond $p_{\mathrm{thr}}=1.05$, as much of the execution time is spent outside the optimized kernels.

\section{Correctness and Failure Breakdown}
\label{app_sec:correctness_failure}
\label{app_subsec:failure_breakdown}

\Cref{tab:cf_breakdown,tab:rf_breakdown,tab:ce_breakdown} provide detailed subcategory counts and percentages for compilation failures (CF), runtime failures (RF), and correctness errors (CE), respectively.

\begin{table}[H]
\centering
\caption{Compilation failure (CF) sub-category breakdown.
\textbf{Syntax Error}: code fails to parse due to malformed statements, incorrect pragma usage, or structural issues.
\textbf{Type/Decl Error}: undeclared identifiers, type conflicts, missing headers, or redefinitions.
\textbf{Malformed Output}: LLM output contains raw delimiters or natural-language text instead of source code.}\label{tab:cf_breakdown}
\small
\begin{tabular}{c r r r r}
\toprule
 & \textbf{EX1} & \textbf{EX2} & \textbf{EX3} & \textit{Post-proc?} \\
\midrule
\textbf{Syntax Error}     & 212 (85.2\%) & 462 (80.5\%) & 124 (53.4\%) & No      \\
\textbf{Type/Decl Error}  & 17 (6.8\%)   & 92 (16.0\%)  & 29 (12.5\%)  & Partial \\
\textbf{Malformed Output} & 20 (8.0\%)   & 20 (3.5\%)   & 79 (34.1\%)  & Yes     \\
\midrule
\textbf{Total}            & 249          & 574          & 232          &         \\
\bottomrule
\end{tabular}
\end{table}

\vspace{1cm}

\begin{table}[H]
\centering
\caption{Runtime failure (RF) sub-category breakdown.
\textbf{Memory Error}: illegal memory access or host-side heap corruption.
\textbf{CUDA Error}: CUDA runtime API errors or other GPU-side execution failures.}\label{tab:rf_breakdown}
\small
\begin{tabular}{c r r r r}
\toprule
 & \textbf{EX1} & \textbf{EX2} & \textbf{EX3} & \textit{Post-proc?} \\
\midrule
\textbf{Memory Error} & 50 (100.0\%) & 13 (100.0\%) & 1 (3.1\%)    & No \\
\textbf{CUDA Error}   & 0 (0.0\%)    & 0 (0.0\%)    & 31 (96.9\%)  & No \\
\midrule
\textbf{Total}        & 50           & 13           & 32           &    \\
\bottomrule
\end{tabular}
\end{table}

\vspace{1cm}

\begin{table}[H]
\centering
\caption{Correctness error (CE) sub-category breakdown.
\textbf{Incorrect Output}: program executes but produces results that differ from the baseline reference.
\textbf{No Output}: program produces no output file or output cannot be parsed for comparison.}\label{tab:ce_breakdown}
\small
\begin{tabular}{c r r r r}
\toprule
 & \textbf{EX1} & \textbf{EX2} & \textbf{EX3} & \textit{Post-proc?} \\
\midrule
\textbf{Incorrect Output} & 268 (79.1\%) & 412 (86.4\%) & 140 (100.0\%) & No \\
\textbf{No Output}        & 71 (20.9\%)  & 65 (13.6\%)  & 0 (0.0\%)     & No \\
\midrule
\textbf{Total}            & 339          & 477          & 140           &    \\
\bottomrule
\end{tabular}
\end{table}

\section{API Cost}
\label{app_sec:budget}

A single full evaluation pass is $92$ benchmarks $\times\,N{=}10$ samples $=920$ generation calls per (model, EX) for EX1 and EX2, and $22\times 10=220$ calls for EX3.
The API cost for GPT-5.1, Claude Sonnet 4.5, and Qwen3-Coder-480B-A35B-Instruct is approximately \$23.5 / \$33.1 / \$8.9 for EX1, \$24.6 / \$33.7 / \$9.1 for EX2, and \$6.5 / \$8.3 / \$2.2 for EX3.

The total API cost across all three settings and models is approximately \$$150$.

\section{License for Existing Assets}
\label{app_sec:license}

\HPCBench draws kernels from existing HPC benchmark suites and evaluates optimizations generated by LLMs accessed through their official APIs.
External profiling tools are used to identify computational hotspots.
\Cref{tab:asset-licenses} summarizes the source, version, license, and URL of the third-party benchmark assets used in this work.
The released artifact contains benchmark metadata, kernel annotations, prompt templates, build scripts, and evaluation code; 
upstream suites are referenced at their original URLs rather than redistributed in full.

\begin{table}[!htbp]
\centering
\small
\caption{Licenses of existing benchmark assets used in \HPCBench.}
\label{tab:asset-licenses}
\begin{tabular}{p{3.0cm} p{2.0cm} p{4.0cm} p{4.6cm}}
\toprule
\textbf{Asset} & \textbf{Version} & \textbf{License} & \textbf{URL} \\
\midrule
\multicolumn{4}{l}{\textit{Benchmark suites}} \\
\midrule
Rodinia~\cite{rodinia}      & 3.1   & University of Virginia BSD-style permissive license & \url{https://github.com/JuliaParallel/rodinia} \\
PolyBench/C~\cite{polybench}  & 4.2.1 & Ohio State University Software Distribution License & \url{https://github.com/MatthiasJReisinger/PolyBenchC-4.2.1} \\
MiBench~\cite{mibench}  & 1.0   & GPL-3.0, for the reference GitHub distribution & \url{https://github.com/ekut-es/mibench} \\
MachSuite~\cite{machsuite} & 1.0 & BSD-compatible; per-file licenses include BSD-3 and ISC & \url{https://github.com/breagen/MachSuite} \\
PARSEC~\cite{parsec}     & 3.0   & Princeton University BSD-3-Clause-style permissive license & \url{https://github.com/bamos/parsec-benchmark} \\
Parboil~\cite{parboil} & 2.5   & University of Illinois/NCSA Open Source License & \url{https://github.com/yuhc/gpu-parboil/} \\
Codee performance-demos~\cite{codee_benchmark} & main (commit-pinned) & Redistributed under explicit permission granted through a prior research collaboration with Codee. & \url{https://github.com/codee-com/performance-demos} \\
\bottomrule
\end{tabular}
\end{table}

\paragraph{Compliance.} 
For licensed upstream benchmark suites, we retain the original copyright headers, license notices, and non-endorsement terms required by the corresponding licenses.
We also cite the original papers or project pages where available.
The released artifact is available at \url{https://github.com/Deep-Learning-Profiling-Tools/HPC-Bench}.

\HPCBenchOrigSection{Limitations and Scope}
\label{app_sec:limitations}

The evaluation has the following limitations:
\begin{itemize}
\item \textbf{EX3 coverage is narrower.}
The CUDA benchmark set has $22$ benchmarks, drawn primarily from Rodinia.
Motif-level macro-averages are therefore less stable for EX3 than for EX1/EX2 and are omitted from the main analysis.
\item \textbf{Hardware/software dependence.}
Reported speedups are tied to a fixed AMD EPYC 7543 + NVIDIA A100 stack with GCC 14.2, NVCC 12.4, and \texttt{-O3}.
Absolute numbers may shift across architectures, compiler versions, and runtime settings.
\item \textbf{Model and prompting dependence.}
The evaluation uses three models (GPT-5.1, Claude Sonnet 4.5, Qwen3-Coder-480B) at a fixed sampling budget ($N{=}10$, default $k{=}3$) and fixed prompt templates.
Results may not generalize directly to smaller open models or to substantially different prompting strategies.
\item \textbf{Kernel-time vs.\ total-time gap.}
Kernel-time speedups can substantially exceed total-time speedups when setup, I/O, allocation, or PCIe transfers dominate wall-clock time (\Cref{app_subsec:total_runtime_metrics}); we report kernel time as the primary signal because the LLM only edits the kernel region.
\end{itemize}



\end{document}